\documentclass[aps,twocolumn,showpacs,floatfix,groupedaddress,amsmath,amssymb,pra]{revtex4-1}
\usepackage{graphicx}
\usepackage{multirow}
\usepackage{color}
\usepackage{mathptm}
\usepackage{collref}

\usepackage{array}
\usepackage{hyperref}

\newcommand{\ket}[1]{\lvert #1 \rangle}
\newcommand{\bra}[1]{\langle #1 \rvert}

\newcommand{\sech}{\mathrm{sech}}

\definecolor{ForestGreen}{rgb}{0.0, 0.27, 0.13}

\newcommand{\fref}[1]{Fig.~\ref{#1}}
\newcommand{\eref}[1]{Eq.~\ref{#1}}

\usepackage[utf8]{inputenc}
\usepackage[T1]{fontenc}
\makeatletter
\DeclareFontFamily{OMX}{MnSymbolE}{}
\DeclareSymbolFont{MnLargeSymbols}{OMX}{MnSymbolE}{m}{n}
\SetSymbolFont{MnLargeSymbols}{bold}{OMX}{MnSymbolE}{b}{n}
\DeclareFontShape{OMX}{MnSymbolE}{m}{n}{
    <-6>  MnSymbolE5
   <6-7>  MnSymbolE6
   <7-8>  MnSymbolE7
   <8-9>  MnSymbolE8
   <9-10> MnSymbolE9
  <10-12> MnSymbolE10
  <12->   MnSymbolE12
}{}
\DeclareFontShape{OMX}{MnSymbolE}{b}{n}{
    <-6>  MnSymbolE-Bold5
   <6-7>  MnSymbolE-Bold6
   <7-8>  MnSymbolE-Bold7
   <8-9>  MnSymbolE-Bold8
   <9-10> MnSymbolE-Bold9
  <10-12> MnSymbolE-Bold10
  <12->   MnSymbolE-Bold12
}{}

\let\llangle\@undefined
\let\rrangle\@undefined
\DeclareMathDelimiter{\llangle}{\mathopen}%
                     {MnLargeSymbols}{'164}{MnLargeSymbols}{'164}
\DeclareMathDelimiter{\rrangle}{\mathclose}%
                     {MnLargeSymbols}{'171}{MnLargeSymbols}{'171}
\makeatother

\begin{document}
\title{MERA as a holographic strange correlator}
\author{Nathan A. McMahon$^{1,2}$, Sukhbinder Singh$^{3}$, Gavin K. Brennen $^{1}$}
\affiliation{$^{1}$Center for Engineered Quantum Systems, Dept. of Physics \& Astronomy, Macquarie University, 2109 NSW, Australia}
\affiliation{$^{2}$Center for Engineered Quantum Systems, School of Mathematics and Physics, The University of Queensland, St Lucia, Queensland 4072, Australia}
\affiliation{$^{3}$Max-Planck Institute for Gravitational Physics (Albert Einstein Institute), Potsdam, Germany}

\begin{abstract}
The multi-scale entanglement renormalization ansatz (MERA) is a tensor network that can efficiently parameterize critical ground states on a 1D lattice, and also suggestively implement some aspects of the holographic correspondence of string theory on a lattice. Extending our recent work  [S. Singh, Physical Review D {\bf 97}, 026012 (2018); S. Singh, N. A. McMahon, and G. K. Brennen, Phys. Rev. D {\bf 97}, 026013 (2018)], we show how the MERA representation of a 1D critical ground state---which has long range entanglement---can be viewed as a \textit{strange correlator}: the overlap of a 2D state with short range entanglement and a 2D product state. Strange correlators were recently introduced to map 2D symmetry protected or topologically ordered quantum states to critical systems in one lower dimension. The 2D quantum state dual to the input 1D critical state is obtained by \textit{lifting} the MERA, a procedure which introduces \textit{bulk} quantum degrees of freedom by inserting intertwiner tensors on each bond of the MERA tensor network. We show how this dual 2D bulk state exhibits several features of holography, for example, appearance of horizon-like holographic screens and bulk gauging of global on-site symmetries at the boundary. We also derive a quantum corrected Ryu-Takayanagi formula relating boundary entanglement entropy to bulk geodesic lengths---as measured by bulk  entropy---and numerically test it for ground states of a set of unitary minimal model CFTs, as realized by 1D anyonic Heisenberg models.
\end{abstract}

\maketitle

In recent years, there has been a push to understand the celebrated anti-de Sitter/conformal field theory (AdS/CFT) correspondence \cite{ref:Large_N_Limit, Witten:1998qj}, a concrete realization of the so called holographic principle, from the perspective of quantum information theory. In particular, some aspects of the AdS/CFT have been realized using tensor network descriptions of ground states of critical quantum many body systems.
It was first suggested by Swingle in Ref.~\cite{ref:Swingle} that the multi-scale entanglement renormalization ansatz (MERA) \cite{ER,MERA}, a particular tensor network suited to describing critical ground states, might also be viewed as a spatial slice of a holographic AdS spacetime. Since then several other holographic interpretations have been presented both of the MERA and other related tensor networks, see e.g. Ref.~\cite{holoMERAr1,holoMERAr2,holoMERAr3,holoMERAr4,holoMERAr5,holoMERAr6,holoMERAr7,holoMERAr8,ref:LiftedMERA_1,ref:LiftedMERA_2}. Even in the absence of general consensus yet on how the MERA realizes holography, one basic lesson is more or less apparent: a given MERA representation can be interpreted in dual (even several) ways.

In this paper, we show how the MERA representation of any 1D critical ground state $\ket{\Psi}$ may also be viewed as a \textit{strange correlator}, namely, as an overlap between a 2D quantum state $\ket{\Psi^{(\rm lift)}}$ with short range entanglement and a 2D product state. Strange correlators were first introduced in Ref.~\cite{StrangeCorrelator} to map 2D symmetry protected quantum phases of matter to 1D critical systems. Using the PEPS tensor network Ref.~\cite{StrangeCorrelator1} extended strange correlators to map 2D topologically phases, described by a topological quantum field theory (TQFT), to 1D critical systems, thus also realizing explicitly the TQFT$_{2+1}$/CFT$_{1+1}$ correspondence. Here, we will show that the `dual' 2D quantum state obtained from the MERA---in addition to having short-range correlations---also exhibits a number of features reminiscent of holography. We suggest that the strange-correlator construction could be useful  as a bulk description in a holographic interpretation of the MERA.
To this end, we generalize our recent work presented in Ref.~\cite{ref:LiftedMERA_1,ref:LiftedMERA_2} in an important direction. There we described how the MERA description of $\ket{\Psi}$ can be ``lifted'' to a 2D quantum state $\ket{\Psi^{(\rm lift)}}$. In this lifting construction---which was shown to caricature several features of the AdS/CFT dictionary---two physical degrees of freedom (DOFs) were introduced on each bond of the tensor network, and the MERA was modified by inserting additional `lifting tensors' (one on each bond) that act on the new bond DOFs. 
But the construction mapped a given critical state $\ket{\Psi}$ to many lifted states, each of which generally had different entanglement. This was because a different lifting tensor was chosen for each different choice of `bond' basis in which the MERA tensors are expressed. This is somewhat unsatisfactory, since a change of bond basis in the tensor network still describes the same state $\ket{\Psi}$
\footnote{A basis-dependent boundary to bulk mapping could nonetheless correspond to an alternative interesting holographic scenario e.g. the different bulk states corresponding to different choice of basis may be related together in an interesting way. However, in this paper, we take the view that demanding basis independence is more natural.}.
%
%
In this paper, we present a modified construction where the different lifted states are related to one another only by on-site unitary transformations---thus, they all have the same entanglement properties. This allows us to associate a unique 2D entanglement structure to each 1D critical MERA state, and thus obtain a strict correspondence between the boundary/bulk entanglement properties.
Our key insight, which we elaborate in the paper, is that the lifting tensor must be an interwiner of the group of basis change transformations: the group $U(\chi)$ where $\chi$ determines the size of the MERA tensors \footnote{The bond basis change can more generally be an arbitrary transformation from the general linear group $GL(\chi)$. However, here we will restrict to the subgroup $U(\chi)$ which corresponds to considering MERA tensor networks that are comprised only of isometric tensors.}.


\begin{figure}[t]
\includegraphics[width=\columnwidth]{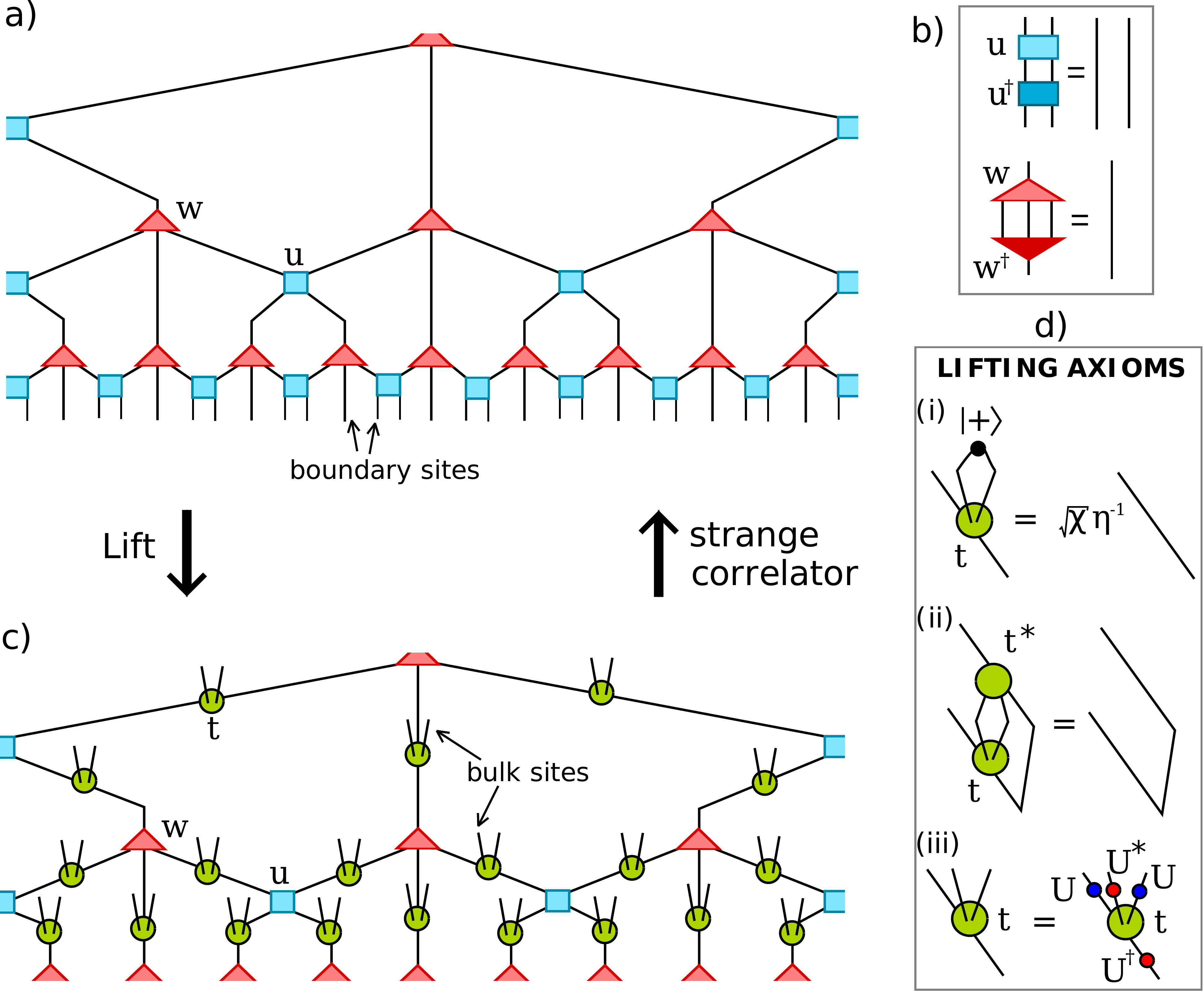}
\caption{(a) A patch of an infinite MERA tensor network, which describes a quantum critical ground state $\ket{\Psi}$ on an infinite lattice $\mathcal{L}$ as follows. Each open index $i$ at the boundary of the tensor network is associated with a site of $\mathcal{L}$ and labels an orthonormal basis $\ket{i}$ on the site. For a given basis state of the lattice, the open indices are fixed to the corresponding values, which yields a closed tensor network. The latter can be contracted to obtain a complex number, which is the amplitude of the basis state in $\ket{\Psi}$. For simplicity we assume that each index of the tensor network runs over $\chi$ values. The tensor network is made of two types of tensors, illustrated here $u$ and $w$. 
(b) All tensors are isometries and fulfill the constraints shown here. 
(c) The lifted MERA---which describes a 2D quantum state $\ket{\Psi^{(\rm lift)}}$---is obtained by inserting a 4-index ($\chi \times \chi \times \chi \times \chi$) lifting tensor $t$ on each bond of the MERA.
(d) The lifting tensor $T$ is required to satisfy the axioms shown described in the text.
}
\label{fig:LiftedMERA}
\end{figure}

Surprisingly, the new basis-independent construction still retains several interesting features of the previous construction in Ref.~\cite{ref:LiftedMERA_1,ref:LiftedMERA_2}, e.g. the promotion of a boundary global onsite symmetry to a local gauge symmetry in the bulk and the appearance of `holographic screens' in the bulk, both of which we revisit in this paper. Additionally, by using entanglement entropy along a geodesic in the bulk as a surrogate for geodesic length we obtain a quantum corrected Ryu-Takayanagi formula  \cite{ref:Ryu-Takayanagi2006,ref:QRT}.
The overall message here is to focus on finding a holographic structure which emerges purely from tensor networks. Our search is guided by the AdS/CFT correspondence, but our results hold independently.


\textit{MERA and lifted MERA.---} We consider a MERA tensor network that defines a class of a quantum many-body states on an infinite one dimensional lattice, see \fref{fig:LiftedMERA}(a). The MERA is particularly well suited to describe critical ground states \cite{MERACFT}. Given a critical Hamiltonian, the approximate MERA representation of its ground state can be obtained e.g. by a variational energy minimization algorithm \cite{MERAAlgo}, and the approximation can be made more accurate by increasing $\chi$ (which increases the number of variational parameters).
Given the MERA representation of a 1D critical ground state $\ket{\Psi}$ we lift it to a 2D quantum state $\ket{\Psi^{(\rm lift)}}$ by inserting a 4-index lifting tensor $t$ on each bond of the MERA, as shown in \fref{fig:LiftedMERA}(c).
We require that the lifting tensor $t$ fufills some reasonable axioms that are depicted in \fref{fig:LiftedMERA}(d). The first axiom allows us to reverse the lifting and recover a properly normalised boundary state:  $\ket{\Psi^{(\rm lift)}}\rightarrow \ket{\Psi}$ by local bulk projectors onto unnormalized singlets $\ket{+}=\sum_{j=1}^{\chi}\ket{j}\ket{j}$, that is
\begin{equation}\label{eq:strange}
\ket{\Psi} = (\bigotimes_{\mbox{\tiny all bulk sites}} \eta \sqrt{\chi}^{-1}\bra{+}) \ket{\Psi^{(\rm lift)}}.
\end{equation}
The second axiom ensures $\ket{\Psi^{(\rm lift)}}$ is a normalised quantum state. It corresponds to demanding that the lifting tensor is isometric 
\footnote{Therefore, just like the MERA, the lifted MERA may also be expressed as a quantum circuit with a bounded-width causal cone. This in particular implies that the expectation value of local observables can be computed efficiently from the lifted MERA, along with being the basis of the holographic screen property.}
. The third axiom, the only axiom that was not assumed in Ref.~\cite{ref:LiftedMERA_1,ref:LiftedMERA_2}, ensures that $\ket{\Psi^{(\rm lift)}}$ is covariant under a change of basis of the original tensor representation of the MERA by a unitary $U$ in the fundamental representation of $U(\chi)$. This axiom implies that the lifting tensor is an intertwiner of $U(\chi)$ which heavily constrains the structure of the lifting tensor to a canonical form (see Appendix \ref{app:BIProof}) 
\begin{equation}\label{eq:liftingtensor}
\vcenter{\hbox{\includegraphics[width=4.5cm]{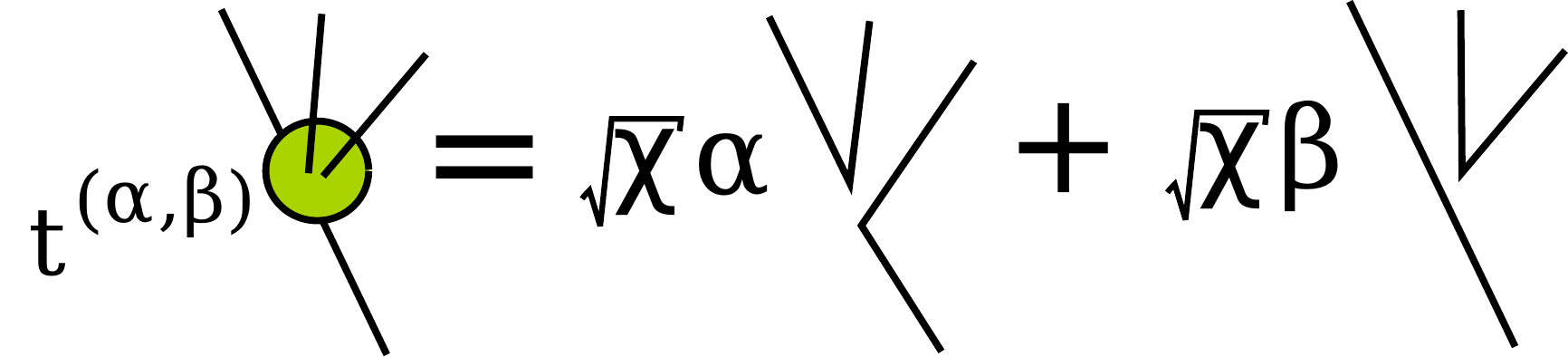}}}.
\end{equation}
Using this solution, the other two lifting axioms imply:
\begin{eqnarray}
\sqrt{\chi}\alpha + (\sqrt{\chi})^{3} \beta = \sqrt{\chi} \eta^{-1},\label{eq:SimpleEquations1}
\\
(|\alpha|^{2} + |\beta|^{2})\chi^{2} + (\alpha \beta^{*} + \alpha^{*}\beta) \chi = 1, \label{eq:SimpleEquations2}
\end{eqnarray}
where $\eta$ is the tuning parameter from the first lifting axiom, \fref{fig:LiftedMERA}(d)(i).
Assuming, without loss of generality, that $\alpha$ and $\beta$ are real and positive yields solutions:
\begin{eqnarray}
\alpha = \sqrt{\frac{1-\eta^{-2}}{\chi^{2}-1}}, \qquad \beta = \frac{\sqrt{\chi^{2}-1}-\eta\sqrt{1-\eta^{-2}}}{\eta\chi\sqrt{\chi^{2}-1}}. \label{eq:SolutionsBI_alpha_beta}
\end{eqnarray}
where $1 \leq \eta \leq \chi$ and $0\leq \alpha,\beta\leq \chi^{-1}$. These values correspond to a legitimate choice $t^{(\alpha,\beta)}$ for the lifting tensor. (Given \eref{eq:SolutionsBI_alpha_beta}, we will also parameterize $t^{(\alpha,\beta)}$ as $t^{(\eta)}$.) A 2D bulk state is defined by choosing a lifting tensor from this domain; different lifting tensors correspond to different bulk states.

\textit{MERA as a strange correlator.---} As introduced in Ref.~\cite{StrangeCorrelator}, a strange correlator is a classical partition function with \textit{algebraically decaying} correlations that is obtained as the overlap of a quantum state with \textit{exponentially decaying} correlations and a product state. We show in Appendix \ref{app:ExpDecayCorr} that the correlation functions in the 2D state $\ket{\Psi^{(\rm lift)}}$ decay exponentially along the geodesic distance between any two bulk sites. This essentially follows from the fact that the tensor network expression for the correlator between any two sites consists of the product of copies of a `transfer matrix' along the shortest bulk path connecting them. By construction, Eq.~\ref{eq:strange}, the critical MERA state $\ket{\Psi}$ is recovered by taking the overlap of the short-range correlated lifted MERA state $\ket{\Psi^{(\rm lift)}}$ with the product state $(\bigotimes_{\mbox{\tiny all bulk sites}} \bra{+})$. Thus, the MERA viewed in this way bears a striking resemblance with a strange correlator. The difference is that a strange correlator is a partition function, a number, while the MERA is a critical quantum state. However, an infinite critical MERA, which we have considered here, can also be understood as a critical partition function: either as a Euclidean path integral on a 2D light-cone geometry of a 1D critical quantum system \cite{ref:Guifre1,ref:Guifre2}; or as a 2D classical partition function $Z_{\rm crit}$, albeit with complex Boltzmann weights  \cite{IBLISDIR2014205}, once we take an additional overlap with a fixed 1D product state $(\bigotimes_{\mbox{\tiny all boundary sites}} \bra{+_x})$, where $\ket{+_x}=\sum_{j=1}^{\chi} \ket{j}$. In this case, we obtain
\begin{equation}
 (\bigotimes_{\mbox{\tiny all boundary sites}}\bra{+_x}) (\bigotimes_{\mbox{\tiny all bulk sites}} \bra{+})  \ket{\Psi^{(\rm lift)}}\propto Z_{\rm crit}.
\end{equation}
In the rest of the paper, we demonstrate that the lifted MERA exhibits features that are reminiscent of holography. For this reason, and based on the above discussion, we additionally qualify the MERA as a \textit{holographic} strange correlator.

\begin{figure}[t]
\includegraphics[width=6cm]{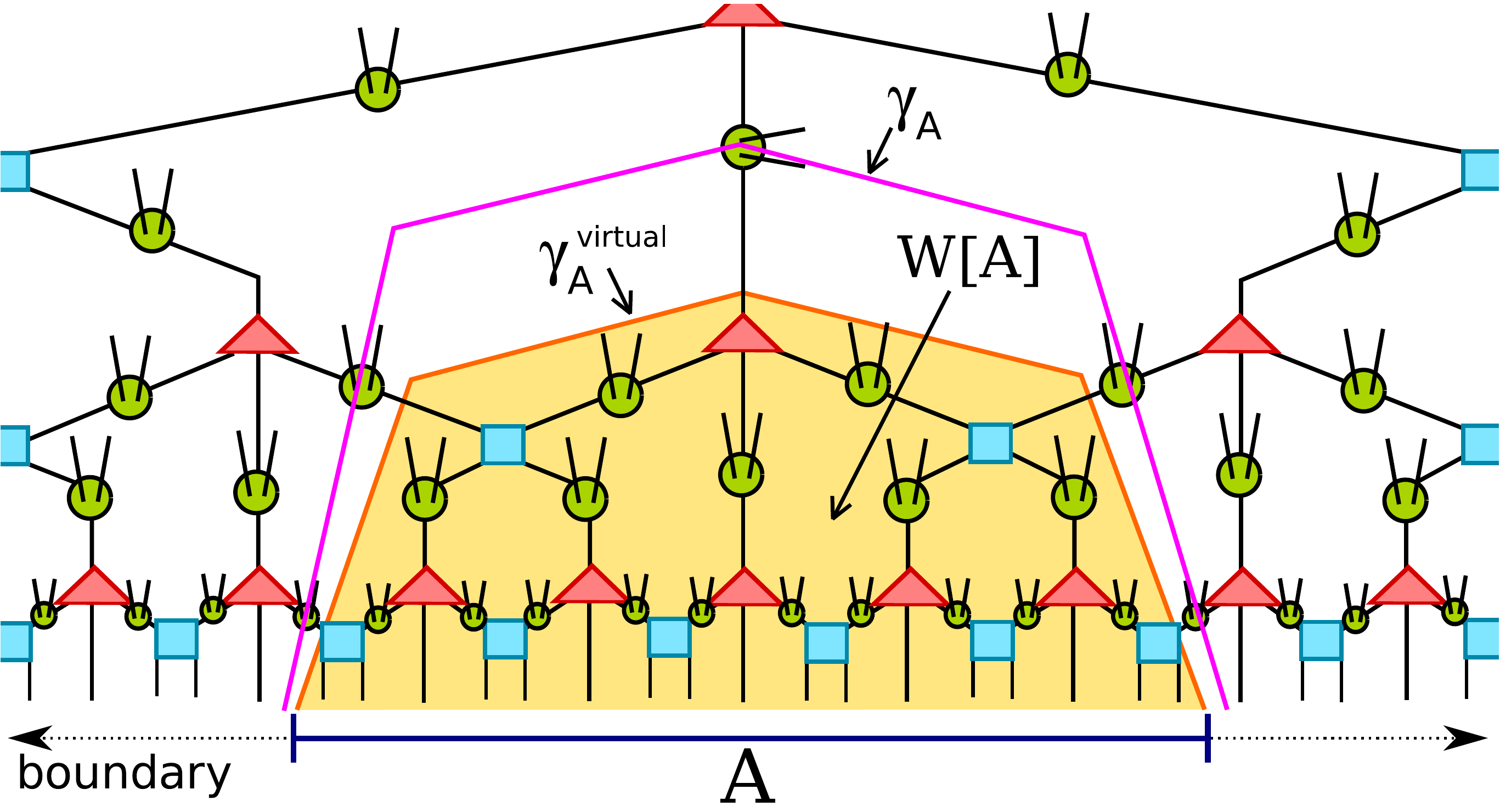}
\caption{The path $\gamma_{A}$ indicates DOFs of a holographic screen on a lifted state $\ket{\Psi^{(\rm lift)}(\eta)}$ that follows a dual geodesic connecting the end points of the boundary region $A$. The bulk DOFs located just outside the 1D path carry all the information contained within the wedge $W[A]$ when $\eta = \eta_{\mathrm{Holo}}$, but are also recovered for any $\eta$ if the corresponding bulk sites along $\gamma_{A}$ are slightly modified by applying a simple filtering operation, as described in Appendix \ref{app:HoloScreen}. On the other hand \emph{virtual} holographic screens, such as $\gamma^{\mathrm{virtual}}_{A}$ exist in the lifted MERA for any value of $\eta$ with no modification. But because these screens are associated with \emph{virtual} DOFs associated with the bonds of the lifted MERA, they are not physically accessible.}
\label{fig:holoscreen}
\end{figure}

\textit{Holographic screens.---} One novel feature of the lifted MERA, first observed in Ref.~\cite{ref:LiftedMERA_1}, is the appearance of holographic screens. A holographic screen is a codimension one surface (the "screen") in the bulk that carries all information contained in the region enclosed between the surface and the boundary. (even if DOFs inside this region are lost, all its information remains intact on the enclosing screen.) Consider, for example, the path $\gamma^{\mathrm{virtual}}_{A}$ shown in \fref{fig:holoscreen}, which encloses the two dimensional wedge $W[A]$. We have:

\begin{equation}
\rho^{\mathrm{bk}}_{W[A]} = R^{\dagger} \rho^{\mathrm{virtual}}_{A}R. 
\label{eq:CutWedgeDef}
\end{equation}

where $\rho^{\mathrm{bk}}_{W[A]}$ is the reduced state of all the bulk sites inside the wedge $W[A]$, $R$ is the tensor obtained by contracting all the tensors inside the wedge, and $\rho^{\mathrm{virtual}}_{A}$ is the reduced density matrix of the \emph{virtual} DOFs associated with the bonds that are intersected by the path $\gamma^{\mathrm{virtual}}_{A}$. Here $R$ is an isometry, namely $R R^{\dagger} = 1_{\chi^{|\gamma_{A}|}}$ (since all the tensors inside the wedge are isometries). \eref{eq:CutWedgeDef} also implies that the traces of moments of the two reduced states are equal; thus there Von Neumann entropies are equal: $S(\rho^{\mathrm{virtual}}_{\gamma_{A}}) = S(\rho^{\mathrm{bk}}_{W[A]})$. While this illustrates how the information in a two dimensional wedge is encoded on a codimension one surface $\rho^{\mathrm{virtual}}_{\gamma_{A}}$, the latter is not physically accessible from the bulk. Thus we refer to $\rho^{\mathrm{virtual}}_{\gamma_{A}}$ as a \emph{virtual} holographic screen.

Remarkably, for the unique value $\eta = \eta_{\mathrm{Holo}} \equiv \sqrt{2\chi}/\sqrt{\chi+1}$, which corresponds to fixing $\alpha = \beta$ in \eref{eq:liftingtensor}, the inaccessible state $\rho^{\mathrm{virtual}}_{\gamma_{A}}$ is exactly equal to the reduced state $\rho^{\mathrm{bk}}_{\gamma_{A}}$ of the \emph{physical} sites located along a path $\gamma_{A}$ which closely follows the virtual screen as shown in \fref{fig:holoscreen}. See Appendix \ref{app:HoloScreen} for details. Thus we have

\begin{eqnarray}
S(\rho^{\mathrm{bk}}_{\gamma_{A}})&=&S(\rho^{\mathrm{virtual}}_{\gamma_{A}})=S(\rho_{W[A]}),
\label{eq:ScreenBoundaryEntropyRelation}
\end{eqnarray}

Furthermore, for any physical local observable $O_{W[A]}$ in $W[A]$ one can determine a \emph{local} observable $O^{\mathrm{bk}}_{\gamma_{A}} \equiv R O_{W[A]} R^{\dagger}$ that clearly has the same expectation value as $O_{W[A]}$ but is supported only on the sites located along $\gamma_{A}$. We refer to $\gamma_{A}$ as the \emph{physical} holographic screen, or simply holographic screen.

More generally, a holographic screen is any path through the lifted network between the end points of a boundary interval that is generated by the greedy algorithm introduced in Appendix \ref{app:HoloScreen}. A holographic screen cuts through only lifting tensors, separating the two DOFs associated with each lifting tensor site; however only the outside DOFs constitute the physical DOFs of the holographic screen $\gamma_{A}$. The greedy algorithm ensures that all the tensors inside the wedge constitute an isometry mapping from $\gamma_{A}^{\mathrm{virtual}}$ to $W[A]$, which leads to the holographic screen property. We remark that the construction of screens via this greedy algorithm and the ability to map bulk operators onto the enclosing screen appears similar to the construction of holographic codes by Harlow et. al. \cite{ref:HaPPY}. 

For given boundary interval A, we can use the greedy algorithm to construct a maximal holographic screen, the screen with the maximal number of tensors in $W[A]$ for some region $A$. We find that the length of the maximal screen can be bounded by a constant multiple of the traditional geodesic path between the end-points of region $A$ (see Appendix \ref{AppRefGeoCone}). Later, we will exploit this property as motivation to assign a physical distance between bulk sites through the maximal holographic screen and to derive a Ryu-Takayanagi like formula.

\begin{figure}[t]
\includegraphics[width=7cm]{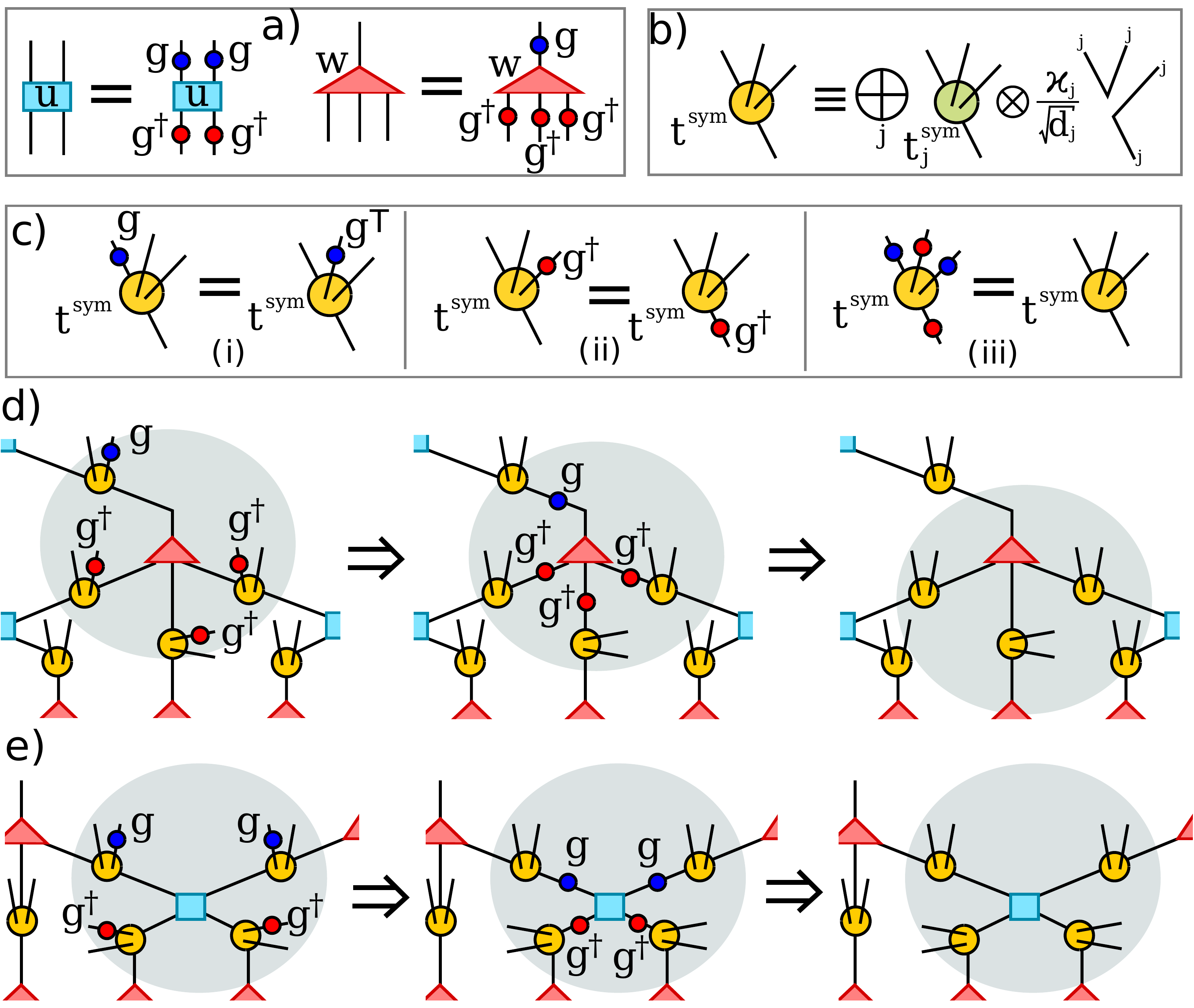}
\caption{a) In a MERA representation of a state that has a global on-site symmetry $\mathcal{G}$, all the tensors can be chosen to commute with the symmetry as shown here; $g$ succinctly denotes a unitary representation of $g \in \mathcal{G}$. 
b) We use the \textit{symmetric} lifting tensor defined as shown here for lifting a symmetric MERA (composed from tensors satisfying panel $a$). Here within a given irrep $j$, $t^{\rm sym}_j$ is a non-symmetric lifting tensor tuned by a parameter $\eta_j$, $\varkappa_j$ is the Frobenius Schur indicator, and $d_j$ is the dim($\mathbb{V}_j$) in \eref{eq:symdecompose}. 
c) Symmetries of the symmetric lifted tensor. 
d) An elementary $w$-gauge transformation---e.g. tensor product of group unitaries acting on bulk sites located immediately around a $w$ tensor---leaves the bulk state invariant as shown. The left and right equalities are obtained by applying panels $c$ and $a$ respectively. 
(e) Similarly, an elementary $u$-gauge transformation leaves the bulk state invariant.}
\label{fig:symLift1}
\end{figure}

\textit{Gauging boundary symmetries.---} In AdS/CFT, a global on-site symmetry in the boundary description generally corresponds to a local gauge symmetry in the bulk. Here we show how our construction can be generalized to implement this feature by introducing a \textit{symmetric} lifting tensor. Our construction follows closely that presented in Ref.~\cite{ref:LiftedMERA_2}, but here we additionally incorporate basis-independence on subspaces that are left unconstrained by the symmetry. Consider that the state $\ket{\Psi}$, which is represented by a MERA, has a global on-site symmetry described by group $\mathcal{G}$, namely, $\ket{\Psi} = (\bigotimes_{s\in \mathcal{L}} U^{[s]}_g) \ket{\Psi}$ for all $g \in \mathcal{G}$ where $U^{[s]}_g$ is a unitary representation of group element $g$ acting on-site $s$ of the lattice and $U^{[s]}_g = U_g$ for all $s$. It turns out that under reasonable assumptions \cite{ref:SymFrac}, if the global on-site symmetry is to be preserved at all renomalization scales than the MERA representation of $\ket{\Psi}$ $\emph{necessarily}$ consists of tensors that commute with $\mathcal{G}$ as depicted in \fref{fig:symLift1}(a), (for sufficiency see also Ref.~\cite{SinghSym,SinghSym1}). It is natural to express the MERA tensors and the lifting tensor in the symmetry basis, in which $U_g$ (or equivalently the vector space $\mathbb{V}$ on which it acts) decomposes as the direct sum of irreducible representations (irreps) as in Ref.~\cite{SinghSym}
\begin{equation}\label{eq:symdecompose}
\mathbb{V} = \bigoplus_j \mathbb{D}_j \otimes \mathbb{V}_j,~~~~~~U_g = \bigoplus_j I_{\mbox{\tiny dim}(\mathbb{D}_j)} \otimes U_{g,j}.
\end{equation}
Here $\mathbb{D}_j$ is the degeneracy space of irrep space $\mathbb{V}_j$, $U_{g,j}$ denotes the unitary corresponding to group element $g$ acting on the irrep space $\mathbb{V}_j$. (Notice that the symmetry acts as the identity $I_{\mbox{\tiny dim}(\mathbb{D}_j)}$ on the degeneracy space.)
 In order to make the symmetry manifest in our construction, we \textit{fix} the symmetry basis $\ket{j,t_j,m_j} = \ket{jt_j} \otimes \ket{jm_j}$ on each bond, where $\ket{jt_j}$ and $\ket{jm_j}$ is a basis in the degeneracy space $\mathbb{D}_j$ and irrep space $\mathbb{V}_j$ respectively. We are still free to choose any basis in the degeneracy spaces since the symmetry acts trivially there. 
In order to generalize our construction to lift a \textit{symmetric} MERA 
\footnote{A symmetric MERA representation of a ground state can be obtained by using the variational energy minimization algorithm adapted to the presence of the symmetry. This algorithm outputs a MERA composed of symmetric tensors and the bond irreps, and their degeneracies, that characterize the ground state, see e.g. Ref.~\cite{SinghSym1}.}
---a MERA composed of tensors that commute with $\mathcal{G}$---we replace the lifting tensor $t$ with the \textit{symmetric lifting tensor} $t^{\rm sym}$ as defined in \fref{fig:symLift1}(b). It can be readily checked that $t^{\rm sym}$ satisfies the symmetries depicted in \fref{fig:symLift1}(c), which in turn imply that the bulk state has a local gauge symmetry---as generated by the elementary gauge transformations depicted in \fref{fig:symLift1}(d,e). When the symmetry group $\mathcal{G}$ is set to identity the symmetric lifting tensor $t^{\rm sym}$ reduces to the non-symmetric version $t$. 

\textit{Ryu-Takanagi formula for bulk/boundary entanglement entropies.---} In the AdS/CFT correspondence, the celebrated Ryu-Takayanagi formula \cite{ref:Ryu-Takayanagi2006} relates the entanglement entropy of a region in the boundary vacuum to the area of the minimal surface that subtends from the region into the bulk. In particular, for 1+1D CFTs, the entanglement entropy of a region $A$ in the vacuum is proportional to the length $L_{\gamma_{A}^{\rm geo}}$ of the geodesic path $\gamma^{\rm geo}_{A}$ between the end points of $A$ through a spatial slice of the dual bulk AdS$_{2+1}$ spacetime: 
\begin{equation}
S (\rho^{\mathrm{CFT}}_A) = \frac{c}{3}\log(|A|) = \frac{L_{\gamma_{A}^{\rm geo}}}{4G^{(2)}}.
\label{eq:Ryu-Takayanagi Formula}
\end{equation}
Here $c$ is the CFT central charge, $G^{(2)}$ is Newton's constant in $2$ space dimensions, and $|A|$ is the length of region $A$ in the flat metric of the boundary CFT. It's important to note that \eref{eq:Ryu-Takayanagi Formula} is the semi-classical Ryu-Takayanagi formula. If instead we have quantum gravity in the bulk \eref{eq:Ryu-Takayanagi Formula} is replaced with $S(\rho^{\mathrm{CFT}}_{A}) = \frac{L_{\gamma_{A}^{\rm geo}}}{4G^{(2)}} + Q$ where $Q$ is the 1-loop additive correction given by the entanglement entropy between the DOFs located inside and outside of the geodesic  \cite{ref:QRT}.

Previous work using unlifted MERA \cite{ref:Swingle} has connected boundary entropy and bulk geodesics via the quantity $|\gamma_{A}|\log(\chi) $, which depends exclusively on the numerical parameter $\chi$. However, from the result of Brown and Henneaux \cite{brown1986}, the radius of curvature of semiclassical AdS$_3$ space is proportional to the central charge according to $c=3R/2G^{(2)}$ and so we would expect the geodesic lengths to vary with theoretical quantity $c$ rather than parameter $\chi$.
We now derive a formula analogous to the quantum-corrected Ryu-Takayanagi formula using our lifting construction. First recall that when $\eta = \eta_{\mathrm{Holo}}$ we find $S(\rho^{\mathrm{bk}}_{\gamma_{A}}) = S(\rho^{\mathrm{bk}}_{W[A]})$. Therefore we define:
\begin{equation}
\frac{\ell_{\gamma_{A}}}{4G^{(2)}} \equiv S(\rho^{\mathrm{bk}}_{\gamma_{A}}).
\label{eq:Ryu-Takayanagi Formula}
\end{equation}
For this reason the function $\ell_{\gamma_{A}}$, a measure of entanglement entropy, is a bonafied measure of length as it is positive, symmetric in boundary points, and satisfies the triangle inequality. See Appendix \ref{App:TriangleInequality} for proof. 

Next consider the state $\ket{\Psi^{(\rm lift)}(\eta=1)}$, in which the bulk DOFs are completely decoupled from the boundary. Thus $S(\rho^{\mathrm{bk}}_{W[A]}(\eta=1))=S(\rho^{\mathrm{CFT}}_A)$, where $S(\rho^{\mathrm{CFT}}_{A})$ is the entanglement entropy of the boundary CFT. Using this fact and extending the definition \eref{eq:Ryu-Takayanagi Formula} away from $\eta = \eta_{\mathrm{Holo}}$ (see Appendix \ref{app:HoloScreen}) we find
\begin{equation}
S(\rho^{\mathrm{CFT}}_A)= \frac{\ell_{\gamma_{A}}}{4G^{(2)}} - Q (W[A]),
\label{eq:Quantum_RT}
\end{equation}
where the subtracted term is
\begin{equation}
Q(W[A]) =  S(\rho^{\mathrm{bk}}_{W[A]})-S(\rho^{\mathrm{bk}}_{W[A]},\eta=1)  > 0.
\end{equation}
To wit, Eq.~\ref{eq:Quantum_RT} equates the entropy of the boundary CFT to an entropic property of DOFs along the geodesic of the bulk quantum state $\ket{\Psi^{(\rm lift)}(\eta=\eta_{\rm Holo})}$ minus a correction term $Q(W[A])$ corresponding to the additional entanglement between the wedge $W[A]$ and the rest of the bulk state. In fact, Eq. \ref{eq:Quantum_RT} can be generalized to all valid lifted states $\ket{\Psi^{(\rm lift)}(\eta)}$ but it is necessary to apply a pre-filtering operation along the screen, which is an identity operation in the special case $\eta=\eta_{\rm Holo}$ (see Appendix \ref{app:HoloScreen})).

The above discussion holds for the non-symmetric lifted MERA. When using symmetric lifting tensors, there is additional entanglement between bulk sites within and without the wedge due to coupling between charge DOFs. To better understand the \eref{eq:Quantum_RT} in the symmetric ase we consider the MERA representation of ground states of unitary minimal model CFTs,
realised in anyonic Heisenberg models \cite{anyonicHam}. (See also Appendix \ref{app:Numerics}.) 
Each such CFT is specified by an integer $k\geq 2$ and an associated Hamiltonian, $H(k)$, acting on a chain of non-Abelian anyons. The anyons are spin$-1/2$ irreps of the quantum group $SU(2)_{k}$, which is a deformation of the usual $SU(2)$ group such that there are no spin projection quantum numbers associated to the anyons and total angular momentum is truncated at $k/2$. The Hamiltonian is: 
\begin{equation}
H(k) = - \sum_{i} h_{i}, \qquad h_{i} = \vcenter{\hbox{\includegraphics[scale=0.4]{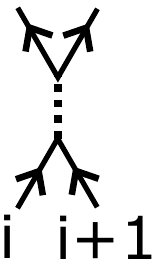}}},
\label{eq:AnyonicHam}
\end{equation}
where the $h_{i}$ term (depicted here in the anyon fusion basis) projects onto the state $\ket{\frac{1}{2}_{i}\times \frac{1}{2}_{i+1} \rightarrow 0}$, i.e. physically the projection onto the spin $0$ fusion space of two (deformed) spin $1/2$ particles at sites $i$ and $i+1$. The Hamiltonian is described by a unitary minimal model CFT with central charge $c(k) = 1-\frac{6}{(k+1)(k+2)}$ \cite{MCAnyonicEnergy} and in the limit $k \rightarrow \infty$ the deformation disappears and the model becomes the bosonic CFT from the usual $SU(2)$ symmetric antiferromagnetic spin $\frac{1}{2}$ Heisenberg model.
\begin{figure}[t]
\includegraphics[width=7cm]{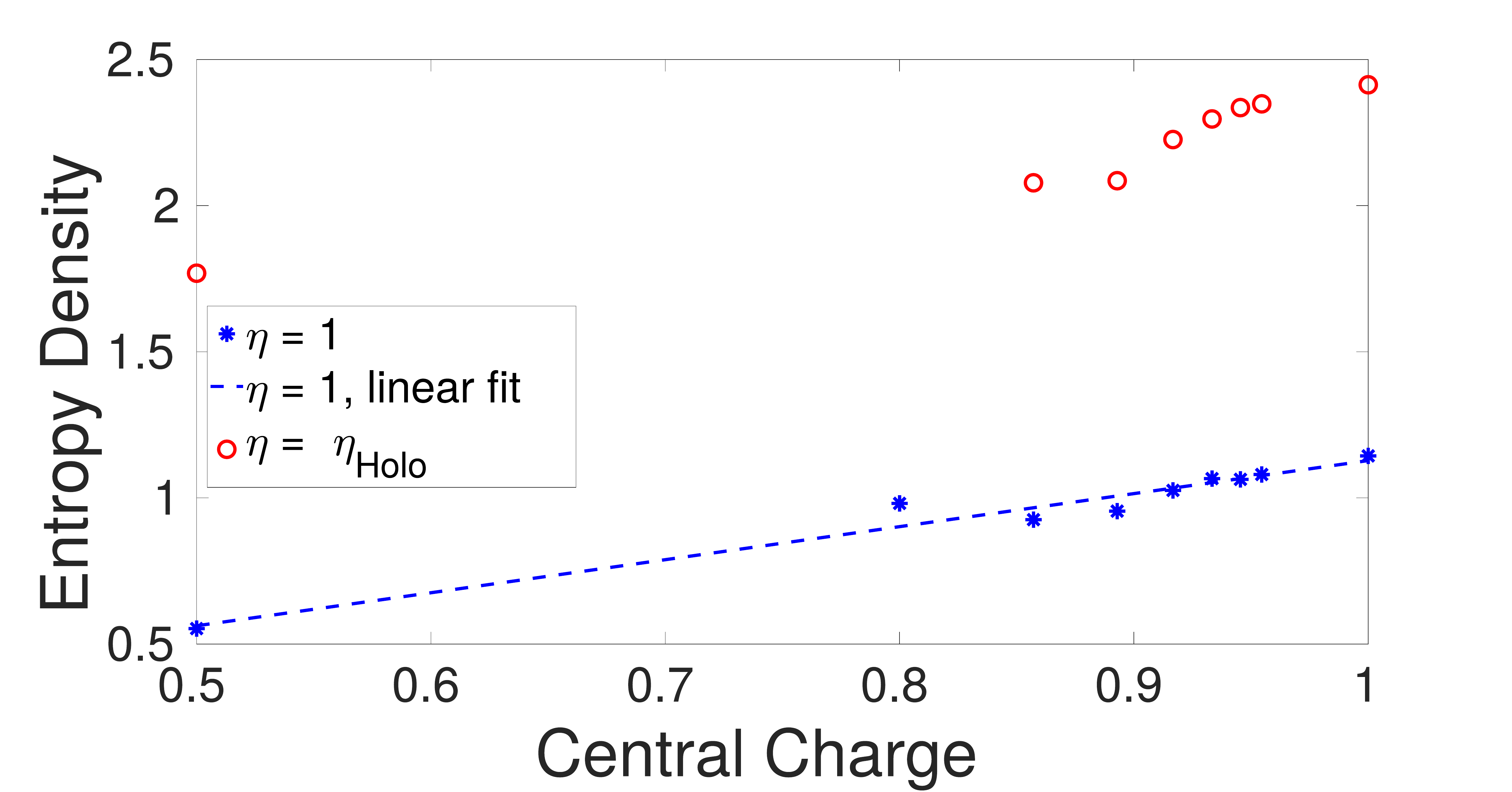}
\caption{
The second R\'enyi entropy density $S^{(2)}({\rho^{\mathrm{bk}}_{\gamma_{A}}}(\eta))/|\gamma_{A}|$ of bulk sites on the geodesic holographic screen $\gamma_{A}$ (see \fref{fig:holoscreen}) for ground states of the Hamiltonian $H(k)$. This is done for $\eta_{j} = 1$ and $\eta_{j} = \sqrt{2\chi_{j}}/\sqrt{\chi_{j}+1}$ (holographic for degeneracy dimension $\chi_{j}$), other values of $\eta$ are included in Appendix \ref{app:Numerics}. We also included the linear regression for the $\eta = 1$ case, see Appendix \ref{app:Numerics} for a discussion of why this relationship should be linear. We have excluded the $k = 3$ points due to lack of numerical convergence and the $k=4$ point for $\eta = \eta_{\mathrm{Holo}}$ due to its anomolous behaviour, see Appendix \ref{app:Numerics} for a deeper analysis of this point.}
\label{fig:NumericalResults}
\end{figure}

We considered Hamiltonians with the values of $k=2,4,5,\ldots,10$ and also $k=\infty$ \footnote{We have excluded $k=3$ due to convergence issues when calculating it, see Appendix \ref{app:Numerics}.}. For each of these, we obtained the MERA representation of the ground state by using the anyonic version of the MERA energy minimization algorithm \cite{anyonicMERA} implimented with Ref.~\cite{ref:SymLibrary}. We then lifted each MERA representation by using the symmetric lifting tensor defined in \fref{fig:symLift1}(b)---where $j$ now labels anyon charges---to obtain the dual bulk states. For each of these, we computed the R\'eyni-2 entropy, $S^{(2)}(\rho) = -\log\left(\mathrm{Tr}\left[\rho^{2}\right]\right)$\footnote{We use the Reyni-2 entropy as it is easier to compute than the von-Neumann entropy for the lifted MERA.}, along the holographic screen illustrated in \fref{fig:holoscreen}. Details about our numerical simulations are described in Appendix \ref{app:Numerics}. \fref{fig:NumericalResults} shows a plot of this bulk entanglement entropy density\footnote{This density is taken with respect to the number of bulk sites in the maximal holographic screen (the path taken through the bulk) as opposed to the boundary size} vs the central charge of the boundary state.
These results demonstrate a linear dependence of bulk entropy density on central charge $c$ for $\eta_{j} = 1$, and passes through 0 at $c = 0$. However the computed slope of $1.131$ is roughly 6 times higher then expected, this increased entropy density can be attributed to charge DOFs meaning that bulk sites are not decoupled from the original MERA. Therefore we find $S(\rho^{\mathrm{bk}}_{W[A]})$ is greater than $S(\rho_{A}^{\rm CFT})$, which would be an interesting prediction to test in other examples of the Ryu-Takayanagi formula. For $\eta=\eta_{\rm Holo}$, setting $\eta_{j} = \sqrt{2\chi_{j}}/\sqrt{\chi_{j}+1}$ for each charge sector, the entropy density tends to grow with the central charge similar to the $\eta = 1$ case
\footnote{We have excluded the $k=4$ point for $\eta_{j} = \eta_{\mathrm{Holo},j}$ here as the lifting procedure for $\eta = \eta_{\mathrm{Holo}}$ appears to amplify numerical instabilities. We have included the full dataset in Appendix \ref{app:Numerics} which includes the $k=4$ point for a variety of $\eta$ values. We also exclude the $k=3$ point since it did not appear to numerically converge to a particular value in our calculations, this point is also excluded from the dataset given in Appendix \ref{app:Numerics}.}
 We also note that there appears to be a constant offset for $\eta_{j} = \eta_{\mathrm{Holo},j}$ (and for other values of $\eta$), for which the entropy density monotonically increase with $\eta$. But since the entropy density at $\eta_{j} = \chi_{j}$ is fixed by the numerical parameter $\chi_{j}$ we expect the bond dimensions to contribute to the entropy density with similar importance to the underlying boundary model from which that the lifted MERA arises, understanding these contributions for both non-symmetric and symmetric models would be useful for future work. These results suggest that even for the symmetric lifted MERA we can interpret the shifts in entropy density for lifted MERA states with  $\eta_{j}>1$ arising from additional entanglement between DOFs from inside and outside the wedge.

\textit{Summary.---} In this work we constructed a short-range entangled 2D bulk state dual to any 1D critical ground state that is described by a MERA tensor network. The critical state is recovered from the 2D bulk state as a strange correlator. We also showed that the 2D bulk state---which is obtained by `lifting' the MERA---exhibits several holographic features: (i) the appearance of holographic screens, (ii) the gauging of global boundary symmetries, and (iii) an analog of the quantum corrected Ryu-Takanayagi formula. A key aspect of this construction is that it yields a unique bulk state (up to on-site unitary transformations) for a given MERA state, which allows a strict correspondence between the entanglement properties of the boundary and the bulk. In particular, we exploited this to use the bulk entanglement entropy as a measure of geodesic lengths, which when compared with the boundary entropy led to a Ryu-Takayanagi like formula. More broadly, our work illustrates a possible way to build a holographic description of the MERA from ground up, by only assuming a reasonable set of input conditions (the lifting axioms and gauging of boundary symmetries). An interesting open question is: under what conditions does the lifted (symmetric) MERA, dual to a critical MERA, describe a state with topological order? If this is possible, then our strange correlator construction could yield a TQFT$_{2+1}$/CFT$_{1+1}$ correspondence---similar to the topological PEPS-based construction presented in Ref.~\cite{StrangeCorrelator1}---but approached from the CFT side.

\begin{acknowledgments}
\textit{Acknowledgements.---}We thank Frank Verstraete for pointing out that the lifting construction presented in this paper (and in previous work \cite{ref:LiftedMERA_1,ref:LiftedMERA_2}) in fact yields a strange correlator. This research was funded in part by the Australian Research Council Centre of Excellence for Engineered Quantum Systems (Project number CE170100009). SS acknowledges the support provided by the Alexander von Humboldt Foundation and the Federal Ministry for Education and Research through the Sofja Kovalevskaja Award. GKB acknowledges support from the ARC through the grant DP160102426.
\end{acknowledgments}


\bibliography{LiftedMERA}

\clearpage

\section{Proof of canonical form for the lifting tensor}\label{app:BIProof}
In this appendix, we prove that \eref{eq:liftingtensor} is the only solution to the third axiom (basis independence) depicted from \fref{fig:LiftedMERA}(d) here:
\begin{equation}\label{eq:Uintertwiner}
\vcenter{\hbox{\includegraphics[width=2.5cm]{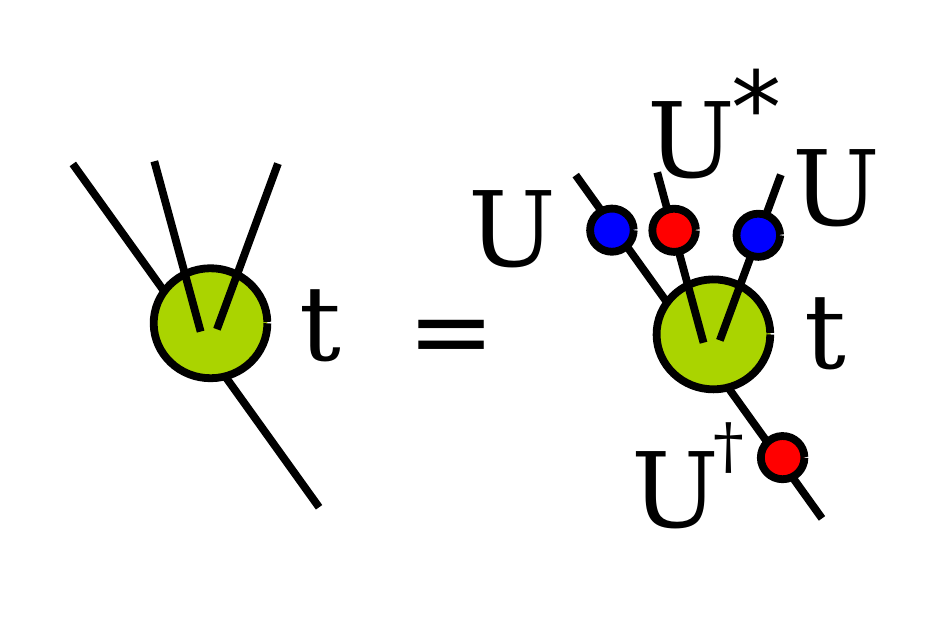}}}
\end{equation}

Here $t$ is the $\chi \times \chi \times \chi \times \chi$ lifting tensor, and the above equation holds for all unitaries $U$ of the group $U(\chi)$ (using the fundamental representation). The group $U(\chi)$ can be decomposed into $U(1)\times SU(\chi)$, that as, as a phase times a special unitary from the group $SU(\chi)$. Since \eref{eq:Uintertwiner} involves two copies of $U$ and two copies of its adjoint (either $U^\dagger$ or $U^{*}$) the phases cancel out. Therefore, we may restrict $U_g$ only to the subgroup $SU(\chi)$. The tensor product (or fusion) of a fundamental and anti-fundamental representations ($f$ and $\bar{f}$ respectively) of $SU(\chi)$ is isomorphic to the direct sum of the trivial representation $0$ and the anti-symmetric representation $A$. \eref{eq:Uintertwiner} then implies that the lifting tensor is an intertwiner of the group $SU(\chi)$. But since we can achieve this by two choices of fusion pathway we obtain the following two possible fusion decompositions (in accordance with the Wigner-Eckart theorem) of the lifting tensor:
\begin{equation}\label{eq:basisproof}
\vcenter{\hbox{\includegraphics[width=\columnwidth]{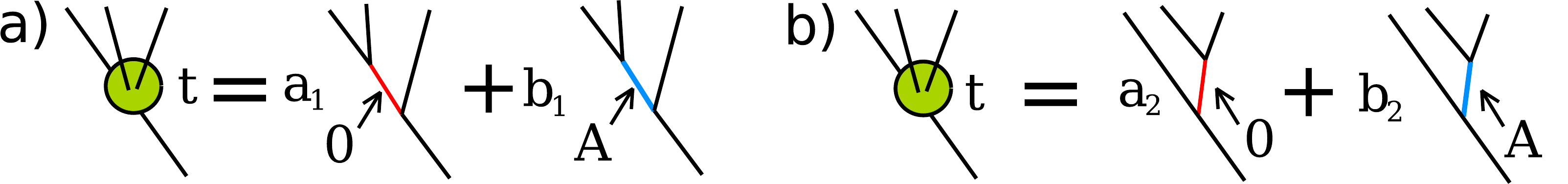}}}
\end{equation}
We also write this as
\begin{eqnarray}
t = a_{1} T_{1} + b_{1} A_{1}\label{eq:Rep1},\\
t = a_{2} T_{2} + b_{2} A_{2}\label{eq:Rep2}.
\end{eqnarray}
where $V_i$ and $A_{i}$ respectively correspond to the diagrams shown in \eref{eq:basisproof}a for $i = 1$, and respectively to the diagrams shown in \eref{eq:basisproof}b for $i = 2$. Specifically, $V_i$ denotes the projection onto the trivial (vacuum) representation, and $A_{i}$ denotes the projection onto the anti-symmetric representation.
Of course since $V_2$ is itself a symmetric $SU(\chi)$ tensor, we can write it as
\begin{equation}
V_{2} = \gamma V_{1} + \delta A_{1} \Rightarrow A_{1} = \delta^{-1} V_{2} - \delta^{-1} \gamma V_{1},
\label{eq:Rewriting}
\end{equation}
where the second equality follows after assuming there is a non-zero contribution of $A_{1}$ to $V_{2}$. If this is not the case and $\delta = 0$ then this indicates that $V_{1} \propto V_{2}$. With the exception of the $\chi = 1$ case (where there is no antisymmetric representation) this is obviously never true.  Lastly we can substitute this into the first representation of $M$ to get a basis in terms of $V_{i}$:
\begin{equation}
M = a_{1} V_{1} + b_{1} A_{1} = a_{1} V_{1} + b_{1} (\delta^{-1} V_{2} - \delta^{-1} \gamma V_{1}) = \tilde{a} V_{1} + \tilde{b} V_{2} \label{eq:SimpleRep}
\end{equation}
The final step, required for completeness, is to show that for any pair $(a,b)$ there exists a pair $(\tilde{a},\tilde{b})$ and vice-versa. We have already shown one direction:
\begin{eqnarray}
\tilde{a} = a + b\delta^{-1} \label{eq:RelateRep1}
\\
\tilde{b} = b \delta^{-1} \gamma \label{eq:RelateRep2}
\end{eqnarray}
Clearly this implies that there is a unique $(\tilde{a},\tilde{b})$ pair for each $(a,b)$ pair. To show the opposite we invert the equations to obtain:
\begin{eqnarray}
a = \tilde{a} - \tilde{b}\gamma^{-1} \label{eq:RelateRepReverse1}
\\
b = \tilde{b} \delta \gamma^{-1} \label{eq:RelateRepReverse2}
\end{eqnarray}
This indicates that if $\gamma \neq 0$ then we also have a unique $(a,b)$ for each $(\tilde{a}, \tilde{b})$. To show this is the case note that if $\gamma = 0$ then we have that tensor $V_{2} \propto A_{1}$. But from \eref{eq:basisproof} it is clear (when bending the top-right most leg down) that $V_{2}$ is the identity on a $\chi^{2}$ dimensional vector space, while $A_{1}$ is a projector onto the $\chi^{2}-1$ dimensional anti-symmetric representation. Therefore, $\gamma \neq 0$.

\section{Bulk state has exponentially decaying correlations}
\label{app:ExpDecayCorr}

\begin{figure}[h!]
\includegraphics[width=7cm]{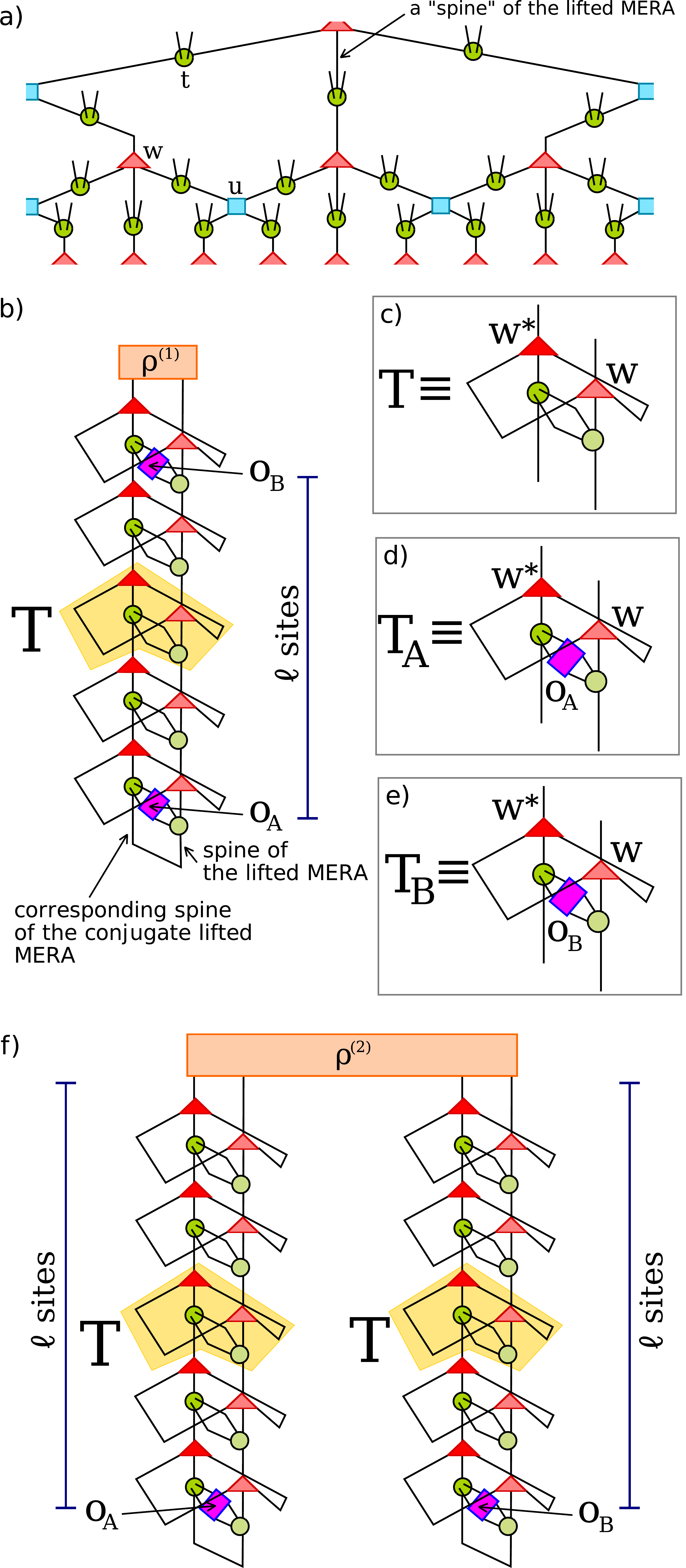}
\caption{(a) A ``spine'' of the lifted MERA, comprised of a 1D chain of $w$-tensors. An infinite number of arbitrarily long spines can be located in the infinite lifted MERA. 
(b) Tensor network expression, \eref{eq:spinecorrelator}, for the 2-point correlator of  two operators $O_A$ and $O_B$ (each acts on a pair of bond sites) that are located deep along the spine. 
(c,d,e) Definitions of $T,T_A,T_B$ that appear in \eref{eq:spinecorrelator}. 
(f) Tensor network expression, \eref{eq:horizcorrelator}, excluding subtraction for local terms $\langle O_{A}\rangle \langle O_{B}\rangle$. In this expression operators $O_A$ and $O_B$ are located at the same depth in the bulk but are now separated along the horizontal direction by a geodesic distance $L = 2\ell$. Here also $O_A$ and $O_B$ are each located at the base of a spine section of the tensor network.}
\label{fig:correlations}
\end{figure}

In this appendix we argue that the bulk states obtained via our lifting construction generally have exponentially decaying correlations. First, consider the bulk correlator $\llangle O_A O_B\rrangle_{\mbox{\tiny spine}}$ of two operators $O_A$ and $O_B$ (each acts on a pair of bond sites) that are located deep in the bulk along the ``spine'' of the lifted MERA, see \fref{fig:correlations}(a). Thanks to the fact that MERA tensors $u,w$ and the lifting tensor $t$ are all isometries, $\llangle O_A O_B\rrangle_{\mbox{\tiny spine}}$ depends only on the tensors that are located along the spine of the lifted MERA (and the corresponding tensors along the spine of the conjugate lifted MERA), while all remaining tensors cancel out, see \fref{fig:correlations}(b). Thus, we obtain the closed-form expression:
\begin{equation}\label{eq:spinecorrelator}
\llangle O_A O_B\rrangle_{\mbox{\tiny spine}} \equiv  \mbox{Tr}\left(\rho^{(1)}  T_A (T)^\ell T_B\right) - \langle O_{A} \rangle \langle O_{B}\rangle
\end{equation}
where $T_A, T^\ell, T_B$ are defined as shown \fref{fig:correlations}(c,d,e), operators $O_A$ and $O_B$ act on bulk sites that are separated by $\ell$ sites, and $\rho^{(1)}$ is the reduced density matrix of one pair of bulk sites on a bond located deep along the spine. For any chosen $w$ the transfer matrix $T$ has dominate eigenvalue $\lambda_{\rm max} = 1$. Therefore, for large $\ell$, $(T)^\ell \approx \ket{\lambda_{\rm max}}\bra{\lambda_{\rm max}}$  is a projector to the dominant eigenvector $\ket{\lambda_{\rm max}}$. This means that the decay of correlations is controlled by the largest eigenvalue less then one, most commonly the second largest eigenvalue $\lambda_2 < 1$, we have $\llangle O_A O_B\rrangle_{\mbox{\tiny spine}} = O(\lambda_2^\ell)$. Thus, correlations decay exponentially along the spine with a correlation length $\xi=1/\ln(\lambda^{-1}_2)$.

Next, consider a 2-point correlator $\llangle O_A O_B\rrangle_{\mbox{\tiny horizontal}}$ of operators $O_A$ and $O_B$ that are located at the same depth in the bulk but are now separated by a distance $L$ along the horizontal direction. Here $L$ is the length of the geodesic between $O_A$ and $O_B$. For simplicity, let us also assume that $O_A$ and $O_B$ are each located at the base of a spine section of the tensor network, and the two spines converge to two neighbouring sites $s,s+1$ as we look deeper into the bulk. Once again, the correlator has a closed-form expression (see \fref{fig:correlations}(f) where $L = 2\ell$):
\begin{equation}\label{eq:horizcorrelator}
\begin{split}
\llangle O_A O_B\rrangle_{\mbox{\tiny horizontal}} \equiv &\mbox{Tr}\left(\rho^{(2)} \left[((T)^\ell T_A) \otimes ((T)^\ell T_B)\right]\right)
\\
& \qquad \qquad - \langle O_{A} \rangle \langle O_{B}\rangle
\end{split}
\end{equation}
where $\rho^{(2)}$ is the reduced density matrix of pair of bulk sites located at neighbouring sites $s$ and $s+1$, and $T$ is the same transfer matrix that appears in \eref{eq:spinecorrelator}. Once again, we find that the correlator $\llangle O_A O_B\rrangle_{\mbox{\tiny horizontal}}$ decays exponentially as $\llangle O_A O_B\rrangle_{\mbox{\tiny horizontal}} = O(\lambda_2^\ell)$.

Here we considered bulk sites located along spines for convenience. The above arguments can be easily generalized for any pair of bulk locations; the closed-form expressions for 2-point correlators away from spines are more complicated and involve also the $u$-tensors. Thus, we find that the bulk correlations decay exponentially in any direction.

\section{Holographic Screen}
\label{app:HoloScreen}

In the main text we pointed out that in our quantum bulk state there are a number of regions of the bulk/boundary of the lifted MERA, $W[A]$, that are equivalent to a surface in the bulk of one dimension lower, the screen $\gamma_{A}$. This relationship is key to our analogous Ryu-Takayanagi formula.
%
%
A key step in this process is to exploit the fact that the lifted MERA is constructed from isometries and unitaries. This then gives rise to a state which we will call $\hat{\rho}_{\gamma_{A}}^{\rm virtual}$, constructed by contracting all tensors outside the virtual screen. This then allows us to connect the states in the wedge $W[A]$, $\rho^{\mathrm{bk}}_{W[A]}$, and $\hat{\rho}_{\gamma_{A}}^{\rm virtual}$ by an isometry $R$ (where $R R^{\dagger} = 1$):

\begin{equation}
\rho^{\mathrm{virtual}}_{\gamma_{A}}= R \rho^{\mathrm{bk}}_{W[A]} R^{\dagger}.
\label{eq:CutWedgeDefApp}
\end{equation}

The intuition behind how the virtual screen works is best illuminated by a constructive greedy algorithm, from which we can construct viable wedges $W[A]$. This greedy algorithm is essentially the same as the greedy algorithm defined in Ref.~\cite{ref:HaPPY}, but slightly restricted here due to us using ordinary isometries rather then perfect or block perfect tensors\cite{ref:HoloCodeBlock}. To perform his algorithm, first choose a boundary $A$, from which we will define a ``wedge'' $W_{0}[A]$ as the empty set of tensors/ the set of physical sites on the boundary $A$. Associated to this is the holographic screen $\gamma_{A,0}^{\rm virtual}$ which is just the boundary $A$. Next define a new wedge $W_{1}[A]$ by choosing a subset of tensors from the original MERA connected to the holographic surface $\gamma_{A,0}^{\rm virtual}$ (the boundary $A$). This subset must include only tensors $u,w: \mathcal{H}_{\rm in} \mapsto \mathcal{H}_{\rm out}$ where the entire space $\mathcal{H}_{\rm in}$ is on the surface $\gamma_{A,0}$, i.e. all downward legs of $u$ and $w$ must be on $\gamma_{A,0}^{\rm virtual}$. This then defines a new wedge $W_{1}[A]$ as the set of tensors $\{T\}$/the boundary sites $A$ and all bulk sites directly below the tensors $\{T\}$. The associated holographic surface $\gamma_{A,1}^{\rm virtual}$ is then the effective boundary for region $A$ generated by excluding said tensors $\{T\}$. This processes can then be repeated any number of times, building $W_{n+1}[A]$ out of $W_{n}[A]$ and $\gamma_{A,n+1}^{\rm virtual}$ out of $\gamma_{A,n}^{\rm virtual}$ by selecting tensors with $\mathcal{H}_{\rm in}$ only along the effective boundary $\gamma_{A,n}^{\rm virtual}$. This generates all holographic surfaces, and we will call the surface associated to the maximal sized set $W[A]$ as the maximal holographic surface. As discussed in Appendix \ref{AppRefGeoCone} this is closely related to the dual graph geodesic. From this construction it is clear that $R$ is an isometry from the effective boundary $\gamma_{A}^{\rm virtual}$ to the true boundary $A$ (and all bulk sites in the wedge $W[A]$.

Reaching equation \eref{eq:CutWedgeDefApp} only requires the MERA begin constructed from isometries and the second lifting axiom (\fref{fig:LiftedMERA}d, ii) via the interpretation of the lifting tensor as an isometry. 

\begin{figure}[t!]
\includegraphics[width=\columnwidth]{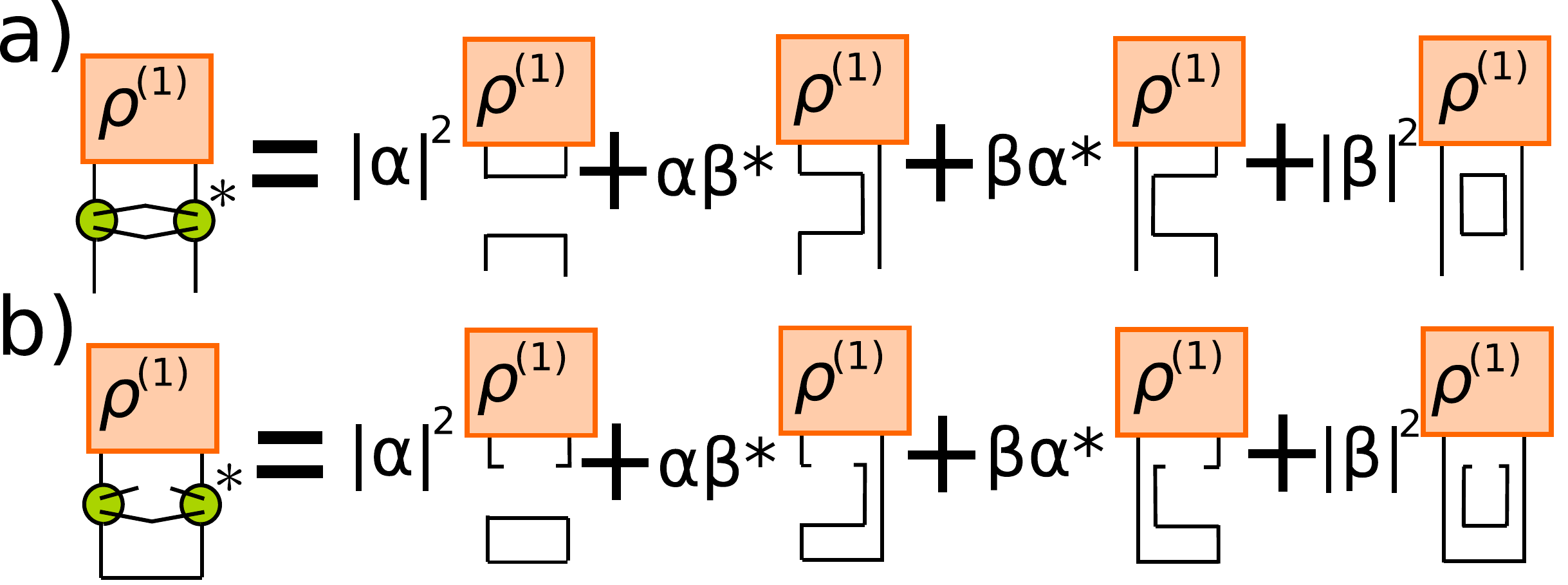}
\caption{An expansion of a single site for states a) $\rho^{\mathrm{virtual}}_{\gamma_{A}}$ and b) $\rho^{\mathrm{bk}}_{\gamma_{A}}$. We contract everything besides these sites into $\rho^{\mathrm{virtual}}_{\gamma_{A}}$ everything else being the same we see that these two states are very similar, but require that we swap $\alpha$ and $\beta$ between diagrams to arrive at an exact equivalence. This diagram is for a single site but can be extended to other multi-site cases by similar equivalences (flipping the $\alpha$ and $\beta$ of each site along the cut one at a time).}
\label{fig:BasisIndependentHolographicScreenExpansion}
\end{figure}

The final step in making the connection between sites in $\rho^{\mathrm{bk}}_{W[A]}$ and on holographic screen $\gamma_{A}$ is to connect the virtual screen to the true holographic screen. This is done by relating the two tensors that appear on the right hand sides of \fref{fig:BasisIndependentHolographicScreenExpansion}(a) and (b). If this can be done then we have said that $\rho^{\mathrm{virtual}}_{\gamma_{A}}$ is equal to the holographic screen $\rho^{\mathrm{bk}}_{\gamma_{A}}$. The support of $\rho^{\mathrm{bk}}_{\gamma_{A}}$ is defined as a subset of the DOFs generated by the lifting tensors along the holographic screen, in particular the DOF deeper into the bulk out of each pair generated by a lifting tensor. Then because $\rho^{\mathrm{virtual}}_{\gamma_{A}}$ is equal to $\rho^{\mathrm{bk}}_{W[A]}$ by an isometry then so is $\rho^{\mathrm{bk}}_{\gamma_{A}}$.

The tensor network described by \fref{fig:BasisIndependentHolographicScreenExpansion}(a) refers to $\rho^{\mathrm{virtual}}_{\gamma_{A}}$ and \fref{fig:BasisIndependentHolographicScreenExpansion}(b) refers to $\rho^{\mathrm{bk}}_{\gamma_{A}}$. Comparing the two right hand sides of \fref{fig:BasisIndependentHolographicScreenExpansion}(a) and (b) it is possible to see that the diagrams associated with coefficient $\alpha \beta^{*}$ (or its complex conjugate) are isotropically equivalent between \fref{fig:BasisIndependentHolographicScreenExpansion}(a) and \fref{fig:BasisIndependentHolographicScreenExpansion}(b). We can also see that the diagrams with coefficients $|\alpha|^{2}$ and $|\beta|^{2}$ are also isotopically equivalent, but swapped in the two diagrams (i.e. the diagram for $|\alpha|^{2}$ in one sum is the same as the diagram for $|\beta|^{2}$ in the other, and vice-versa). Therefore to equate these two diagrams we need to swap $\alpha$ and $\beta$.

To do this we define a completely positive operator $\mathcal{F}$ which acts on all DOFs arising from the lifting tensors along $\gamma_{A}$. It is obvious that when $\alpha = \beta$ then this operator is the identity, earning this limit of $\eta$ its designation as the holographic limit of lifting, $\eta = \eta_{\rm Holo} = \frac{\sqrt{2\chi}}{\sqrt{\chi+1}}$. When we extend $\eta$ away from the holographic limit this $\mathcal{F}$ takes the form of a filtering operation constructed from local operators

\begin{equation}\label{eq:FlipOperatorSingleSite}
\vcenter{\hbox{\includegraphics[width=0.5\columnwidth]{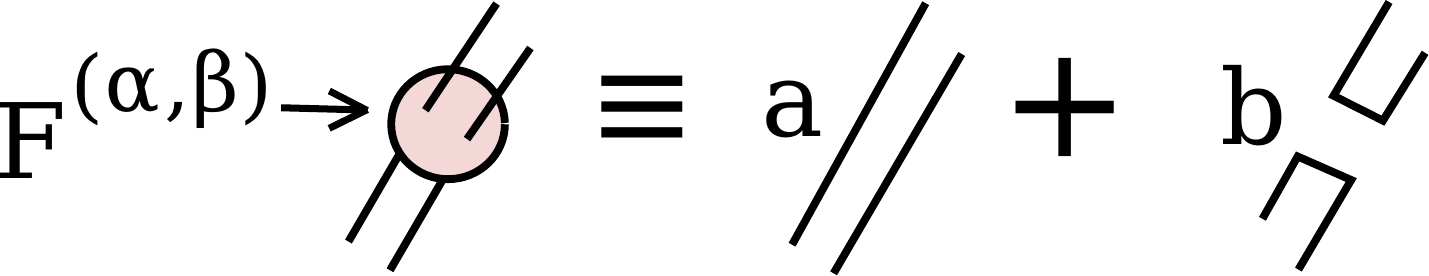}}}.
\end{equation}
Where we set $a = \beta \alpha^{-1}$ and $b = (\alpha^{2} - \beta^{2})\alpha^{-1}(\alpha+\chi \beta)^{-1}$, with these parameters the operator is completely positive. Graphically the action of this on a lifting tensor is

\begin{equation}\label{eq:FlipOperatorAction}
\vcenter{\hbox{\includegraphics[width=0.5\columnwidth]{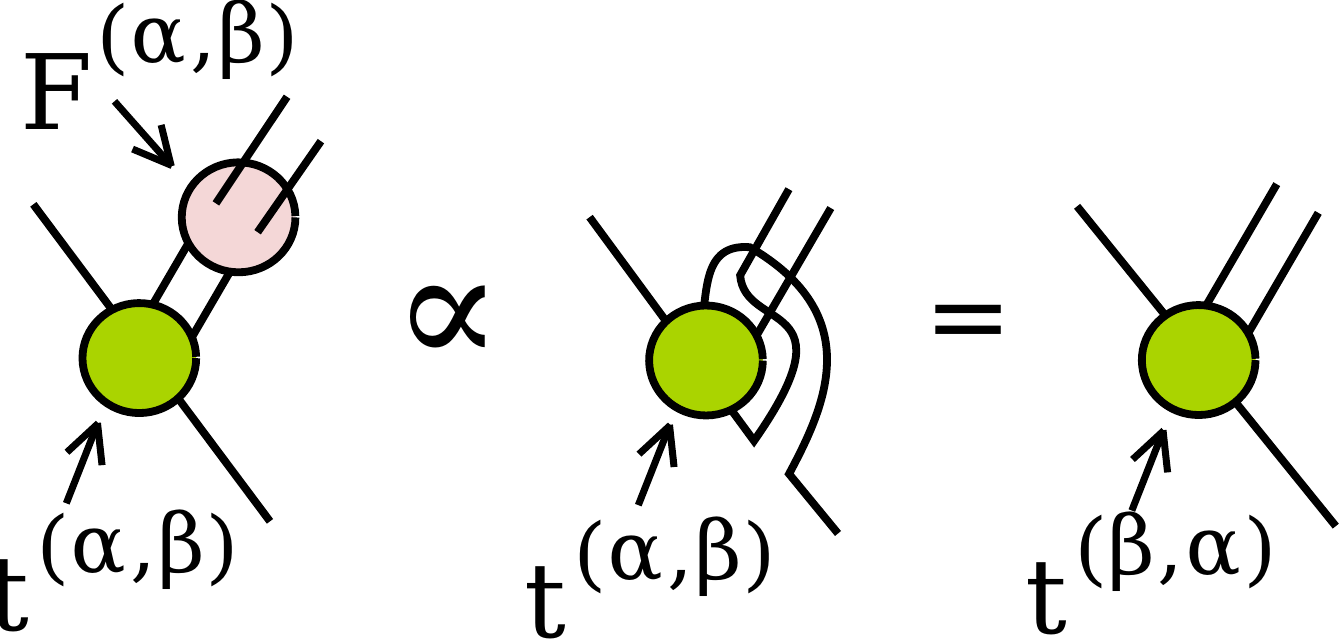}}}.
\end{equation}
where the renormalisation of the CP map, given in \eref{eq:CPmap_1} and \eref{eq:CPmap_2}, is ignored.

Since this operation swaps $\alpha$ and $\beta$ then we can view it as converting $\eta$ to $\tilde{\eta}$:

\begin{equation}
\tilde{\eta}(\eta) = \sqrt{1-\frac{\chi^{2}-1}{\chi^{2}} \left(-\sqrt{\frac{1-\eta^{2}}{\chi^{2}-1}} + \frac{1}{\eta}\right)^{2}}
\end{equation}

We will use $\mathcal{F}$ to indicate the flip of $\alpha$ and $\beta$ for the lifting tensors along the holographic screen sites, therefore we define a modified density matrix $\widetilde{\rho}^{\mathrm{bk}}_{\gamma_{A}}$ with the same support as $\rho^{\mathrm{bk}}_{\gamma_{A}}$ by:

\begin{equation}
\widetilde{\rho}^{\mathrm{bk}}_{\gamma_{A}} = \mathrm{Tr}_{\mathrm{bt}}\left(\mathcal{F}\left[\rho^{\mathrm{bk},(2)}_{\gamma_{A}}\right]\right) = \rho^{\mathrm{virtual}}_{\gamma_{A}}= R \rho^{\mathrm{bk}}_{W[A]} R^{\dagger}
\end{equation}

Here we apply the filtering operation $\mathcal{F}$ to the quantum state $\rho^{\mathrm{bk},(2)}_{\gamma_{A}}$ reduced to only (the pair of) DOFs generated by the lifting tensors along $\gamma_{A}$. Having applied the filter we then trace out the bottom DOF of each lifting tensor along $\gamma_{A}$ by $\mathrm{Tr}_{\mathrm{bt}}$ and reduced the support.
By the previous arguments this is equal to $\rho^{\mathrm{virtual}}_{\gamma_{A}}$ and therefore also equal to $R\rho^{\mathrm{bk}}_{W[A]}R^{\dagger}$. Therefore when extending our Ryu-Takayanagi formula to $\eta$ beyond the holographic limit we replace $S(\rho^{\mathrm{bk}}_{\gamma_{A}})$ with $S(\widetilde{\rho}^{\mathrm{bk}}_{\gamma_{A}})$ in Eq.~\ref{eq:Ryu-Takayanagi Formula}.


Beginning with the completely positive operator from \eref{eq:FlipOperatorSingleSite} from which we constructed the filtering operators, we can construct a local POVM on each site such that one of the measurement outcomes corresponds to successfully flipping $\alpha$ and $\beta$. For $a > |a+b\chi|$ we find the POVM to be:

\begin{equation}\label{eq:CPmap_1}
\begin{split}
M_{1} &= \frac{F^{(\alpha,\beta)}}{a}
\\
M_{2} &= \sqrt{1-\left(\frac{a+b\chi}{a}\right)^{2}} \ \ \mathcal{P}^S
\end{split}
\end{equation}
Where $\mathcal{P}^S$ is the projector onto the singlet. Otherwise we find the POVM to be:

\begin{equation}\label{eq:CPmap_2}
\begin{split}
M_{1} &= \frac{F^{(\alpha,\beta)}}{|a+b\chi|}
\\
M_{2} &= \sqrt{1-\left(\frac{a}{a+b\chi}\right)^{2}} \mathcal{P}^S
\end{split}
\end{equation}

The probability of success for measurement outcome 1 (successfully flipped $\alpha$ and $\beta$) can also be computed making use of the fact that it is acting on a bulk lifted state. Because of that the probability of outcome 1 for local operator at some site can be worked out by realising that the expectation value of the identity is $1$ (as the lifted MERA is a pure quantum state). Further when computing the expectation value of the projector $P_{s} = \frac{1}{\chi} \sum_{j=1}^{\chi} |jj\ket\bra jj|$ on a single site the outcome will be $\eta^{-2}$. This can be observed from the first lifting axiom, which states the expectation value of the bulk state times the singlet is equal to the $\eta^{-1}$ times the norm of a lifted state where we have the decoupled lifting tensor at that site (i.e. the lifting tensor generated by $\eta = 1$). Since this lifting is again a pure state that gives expectation value $\eta^{-2}$. Therefore for a lifted tensor on any single site, the probability of success of the filtering operation is:

\begin{equation}
P=
\Bigg\{
\begin{array}{ccc}
  \frac{a^{2}+(2ba\chi+b^{2}\chi^{2})\eta^{-2}}{a^{2}} & \  a>|a+b\chi|  &  \\
\frac{a^{2}+(2ba\chi+b^{2}\chi^{2})\eta^{-2}}{|a+b\chi|^{2}}  & \  a<|a+b\chi|  &  \\ 
\end{array}
\end{equation}


Finally it is important to note how this changes when we consider symmetries. For abelian symmetries nothing changes except that we define the flip operator separately on each charge $j$. For non-abelian symmetries we choose to flip only the degeneracy DOFs and leave the gauge DOFs alone.

\section{Triangle Inequality of Holographic Cuts}
\label{App:TriangleInequality}

One possible issue with the analogous Ryu-Takayanagi formula is the use of entanglement entropy along holographic screens as the measure of bulk length. In this appendix we demonstrate that this measure of bulk length satisfies the requirements to be a measure of distance for all points on the boundary of the lifted MERA. Furthermore this can be extended to satisfy the requirements of the triangle inequality for a significant class of triangles located fully within the bulk.

This text will focus on the general case of a distance measure between points $a$ and $b$. If $a$ and $b$ are on the boundary then these are points located between physical sites on the boundary, if they are in the bulk then they correspond to plaquettes. The restriction we place on these points is that if either of them are in the bulk then there must exist a holographic screen that passes through both $a$ and $b$, though we will later conjecture an extension to all pairs of $a$ and $b$.

The definition of length that we use is the entropy of the effective boundary $\gamma_{(a,b)}$ associated to the effective boundary sites that exist between $a$ and $b$ along the maximal holographic screen $\gamma_{a,b}$ which passes through these points:
\begin{equation}
\ell_{(a,b)} = S(\rho^{(a,b)}_{\gamma_{a,b}}).
\end{equation}
Where $\rho^{(a,b)}_{\gamma_{a,b}}$ is the reduced density matrix along the holographic screen $\gamma_{a,b}$ connecting $a$ and $b$. 

To be a measure of distance this must satisfy the following four requirements, 1) it must be non-negative $\ell_{(a,b)} \geq 0$, which all entropies satisfy. 2) The distance measure can only be zero if $a = b$, this is true for our measure since a selection of sites at some renormalisation scales in the MERA tensor network (not necessarily critical) is highly unlikely to be a pure state, further if $\eta \neq 1$ then it can never by zero due to a maximally mixed contribution that the lifting tensor directly above the screens will leave. 3) the distance measure must be symmetric so that $\ell_{(a,b)} = \ell_{(b,a)}$, since it is obvious that $\gamma_{a,b} = \gamma_{b,a}$ then this is satisfied also satisfied. 4) Finally the distance measure must satisfy the triangle inequality, for which we have a partial solution which we will spend the rest of this appendix demonstrating.

As mentioned in the main text, the entropy of holographic cuts obeys a triangle inequality when used as a definition of length. As stated, we will restrict to defining the lengths of paths between plaqueettes $a$ and $b$ that lie along a single holographic cut $\gamma_{a,b}$. This is geodesic between $a$ and $b$ if we choose the holographic cut such that this quantity $\ell_{(a,b)} = S(\rho^{(a,b)}_{\gamma_{a,b}})$ is minimised. Since there are possibly multiple holographic screens satisfying this we find that the geodesic $\ell_{(a,b)}$ corresponds to the maximal holographic screen that passes through $a$ and $b$. Considering the entropy along the effective boundaries that make up the screens $\gamma_{(a,b)}$ and $\gamma_{(a,b)}^{'}$ that differ by only one tensor ($\gamma_{(a,b)}$ being more maximal), we find that the entropies are related by:

\begin{equation}
\rho^{(a,b)}_{\gamma_{(a,b)}^{'}} = \varepsilon(V \rho^{(a,b)}_{\gamma_{(a,b)}} V^{\dagger})
\end{equation}

Where $V$ is an isometry of the original MERA and $\varepsilon$ is a local CPTP map which either maximally mixes the site, or leaves it alone. This corresponds to the lifting tensors on the sites at the bottom of $V$ which were part of $\gamma_{(a,b)}$ but are not part of path $\gamma_{(a,b)}^{'}$. This type of CPTP map will \emph{always} increase the entropy so that:

\begin{equation}
\ell_{\gamma_{(a,b)}^{'}} = S(\rho^{(a,b)}_{\gamma_{(a,b)}^{'}}) \geq  S(\rho^{(a,b)}_{\gamma_{(a,b)}}) = \ell_{\gamma_{(a,b)}}.
\end{equation}

Now if we have 3 points $a, b, c$ with holographic screens $\gamma_{a,b}$ and $\gamma_{b,c}$ such that $b$ is between sites $a$ and $c$ in the angular directions, then we can find the holographic screen $\gamma_{a,c}$ which minimises $\ell_{(a,c)} = S(\rho^{(a,c)}_{\gamma_{a,c}})$ (i.e. the "geodesic" path between $a$ and $c$ required to complete the triangle). By the argument above the geodesic screen $\gamma_{(a,c)}$ must be the maximal holographic screen passing through $a$ and $c$. Since the combination of paths $\gamma_{(a,b)}$ and $\gamma_{(b,c)}$ is a holographic screen, then the geodesic holographic screen $\gamma_{(a,c)}$ must be either this screen, or deeper into the bulk. Therefore:

\begin{equation}
S(\rho^{(a,c)}_{\gamma_{a,c}}) \leq  S(\rho^{(a,b,c)}_{\gamma_{a,b}\cup \gamma_{b,c}}) \leq S(\rho^{(a,b)}_{\gamma_{a,b}^{'}}) +  S(\rho^{(b,c)}_{\gamma_{b,c}})
\end{equation}
where the last inequality arises from the subaddativity of entropy and we demonstrate that these lengths satisfy the triangle inequality $\ell_{a,b} + \ell_{b,c} \geq \ell_{a,c}$ for this measure of length in the bulk. 

Therefore this distance measure defines a metric for a large number of sites in the bulk. However some paths do not have a holographic screen between them (for example the faces above and below a disentangler), so here we conjecture a generalisation for distance,
\begin{equation}
\tilde{\ell}_{a,b} = \min_{x} \ell_{a,x} + \ell_{x,b}
\end{equation}
where $x$ is a site sharing (possibly different) holographic cuts with both $a$ and $b$.

The triangle inequality is then
\begin{equation}
\tilde{\ell}_{a,c} = \min_{x} \ell_{a,x} + \ell_{x,c} \leq \min_{y} \ell_{a,y} + \ell_{y,b} + \min_{z} \ell_{b,z} + \ell_{z,c}
\end{equation}

We do not currently know how to prove the triangle inequality for this extension, but already the results here show that we can extend this distance measure into sites within the bulk. As a side note there is a strong relationship between this measure of distance and the geodesics for the dual graph because of the use of maximal holographic screens. As discussed in Appendix \ref{AppRefGeoCone} these maximal holographic screens are related to geodesic screens on the dual graph of the tensor network and are exactly them in the binary and modified binary MERA cases.

\section{Relationship between holographic cut length and geodesic length}
\label{AppRefGeoCone}


In the main text we talked about maximal holographic screens $\gamma_{A}^{\mathrm{screen}}$ of a boundary region $A$ and used them in place of geodesic screens $\gamma_{A}^{\mathrm{geo}}$ when drawing an analogy to the Ryu-Takayanagi formula. In this appendix we will show that the number of sites in the maximal holographic screen for region $A$ is equal to the number number of sites in the geodesic screen connecting the boundaries of region $A$ through the bulk for the binary and modified binary MERAs, while in the ternary case it is strictly bounded above by three times this number:

\begin{equation}\label{eq:HolographicScreenBounds}
|\gamma_{A}^{\mathrm{Geo}}| \leq |\gamma_{A}^{\mathrm{Holo}}| < 3 |\gamma_{A}^{\mathrm{Geo}}|
\end{equation}
and are therefore linear in each other.

The maximal holographic screen, $\gamma_{A,\mathrm{max}}^{\rm screen}$ is unique for a region $A$ and is constructed in a method similar to the greedy algorithm introduced in the HaPPY paper \cite{ref:HaPPY}. This method of construction is to iteratively consider a screen $\gamma_{A,n}^{\rm screen}$ for region $A$. Then for all tensors which have edges passing through $\gamma_{A,n}^{\rm screen}$, if the tensor is an isometry from screen $\gamma_{A,n}^{\rm screen}$ to $(\gamma_{A,n}^{\rm screen})^{'}$ (with the smaller dimensional space on screen $(\gamma_{A,n}^{\rm screen})^{'}$) then we include them in the wedge for screen $\gamma_{A,n+1}^{\rm screen}$. If there are no more tensors to add then we have found the maximal holographic screen $\gamma_{A}^{\mathrm{Holo}} = \gamma_{A, \mathrm{max}}^{\mathrm{screen}}$. With this definition our maximal holographic screen would match the geodesic used for the Ryu-Takayangi formula in the HaPPY holographic codes if we chose to lift tensor networks made out of perfect tensors\cite{ref:HaPPY}.

\begin{figure}[t!]
\includegraphics[scale=0.19]{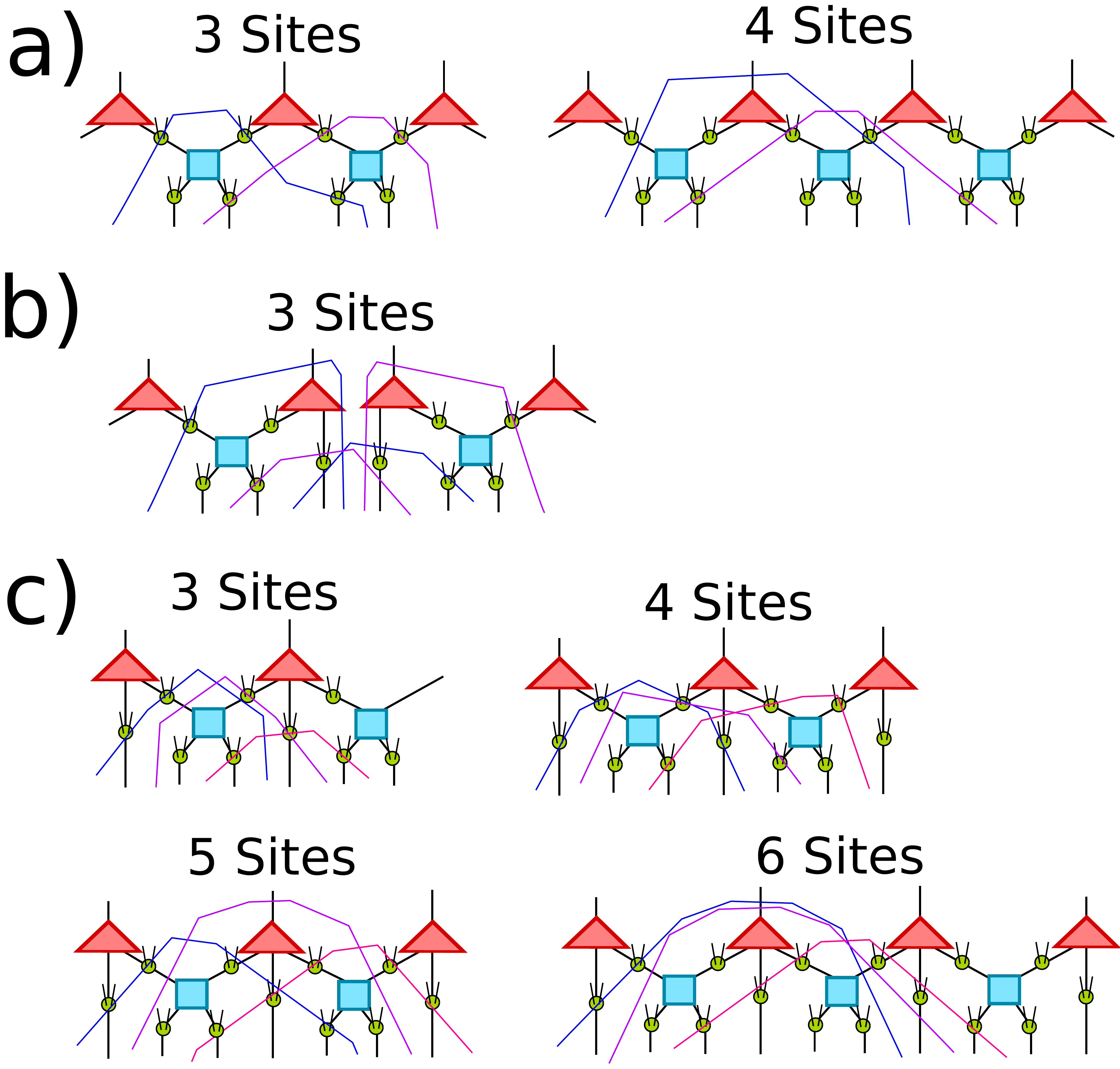}
\caption{Geodesic paths for all possible locations of a variety of sizes for the lifted versions for a (a) binary MERA, (b) modified binary MERA, and (c) ternary MERA. For all 3 MERA cases the paths are only shown for sizes where it is not guaranteed that the geodesic path must pass into the next layer of the MERA. Also in all these cases there can be degeneracies in the geodesic path chosen, here we choose them to all be holographic screens.}
\label{fig:TopBruteForce}
\end{figure}

To compare this to the geodesic path, based on dual graph length, we will separate the tensor network into the different MERA layers an ask if the geodesic connecting edges of a boundary region of length $|A|$ will pass through sites in the above layer. If it would then we can work out the geodesic by considering the the quickest path to the next layer, if not we can just brute force the solution. In general we need to look at the next layer up if within region $A$ of size $|A|$ the algorithm described above would reach the next layer, this happens for $|A|\geq 5$, $|A| \geq 4$ and $|A|\geq 7$ for the binary, modified binary and ternary MERAs respectively. The brute force solutions are all geodesics are shown in \fref{fig:TopBruteForce}, for these we have chosen to show only geodesic cuts which are also holographic cuts, these geodesics are not unique but all cuts in these figures are geodesics.

\begin{figure}[h]
\includegraphics[scale=0.28]{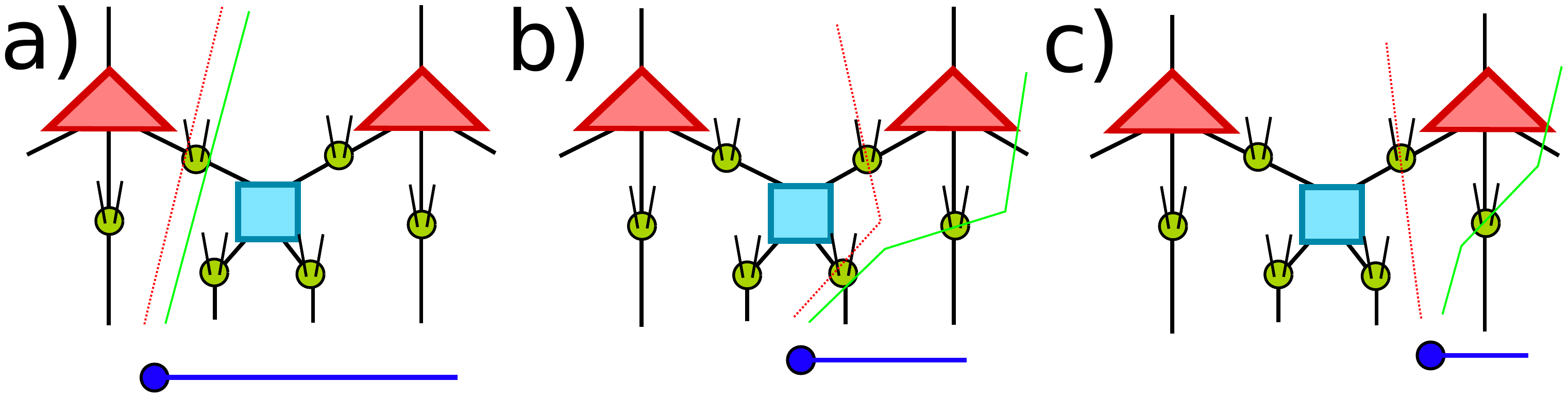}
\caption{A comparison of geodesic paths and holographic paths to reach the next layer of renormalisation for the lifted ternary MERA. Here we are only considering the left-most side of the region of interest, indicated below by the blue line (ending at the blue dot). The argument on the right hand side proceeds in the same manner. If this starts at site (a) then the geodesic path is the same as the holographic path, but for cases where it starts at (b) and (c) these paths differ. In the case where the holographic path differs from the geodesic path the geodesic path ends one site further to the left but performs one less cut then the holographic path does.}
\label{fig:LeftBruteForce}
\end{figure}

For sizes larger then this we try to choose the quickest path to get from our current layer to the next layer up. \fref{fig:LeftBruteForce} shows both the quickest (least cuts) holographic path and quickest (least cuts) geodesic path from one layer to the next, specifically for the left hand side of the region. For the binary and modified binary MERAs these always agree, but for the ternary binary we find that there are two cases where they disagree. This is sufficient to characterise the geodesics for the binary and modified binary MERAs by the maximal holographic screen, finding $|\gamma_{A}^{\mathrm{Holo}}| = |\gamma_{A}^{\mathrm{geo}}|$.

\begin{figure}[h]
\includegraphics[scale=0.18]{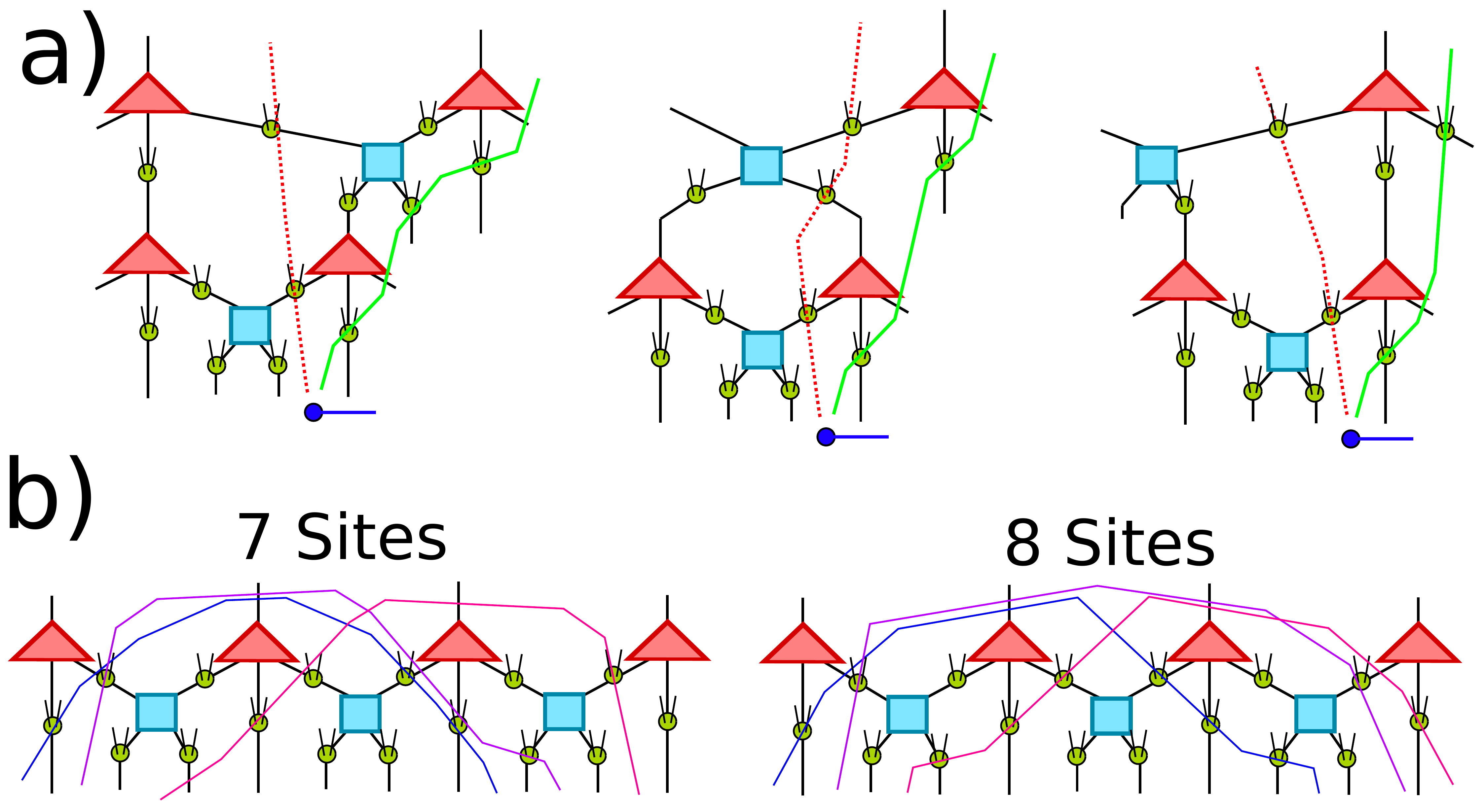}
\caption{a) The three cases for geodesic path compared to holographic path if the geodesic path starts one site to the left (shown by having \fref{fig:LeftBruteForce}c at the bottom). In all cases the geodesic path ends one more site to the left at the next layer but will never cut more bonds the the holographic screen (even though they start at different sites) when getting to the next layer. b) The geodesic paths for 7 and 8 sites for the ternary MERA, all of these paths pass through the next layer up. Like the ones in \fref{fig:TopBruteForce} these geodesic paths are holographic paths, differences between holographic paths and geodesic paths only appear if we pass through at least 2 layers, but that is not needed here to characterise all the holographic and geodesic screens for arbitrary regions of the ternary MERA}
\label{fig:ConcatBruteForce}
\end{figure}

Finally we will finish analysing the ternary MERA, to do this we need to consider the difference in the geodesic path lengths and holographic path lengths on the left side of the region as we go up layers. Here we note that if we go up one layer either the geodesic path and holographic paths are the same or the geodesic path is cuts one fewer edges and is one location to the left of the holographic path at the start of the next layer (see \fref{fig:LeftBruteForce}). We find that on any subsequent layers this difference will always stay at one site by the brute force calculation in \fref{fig:ConcatBruteForce}(a). We repeat this until the effective size of region $A$ at the layer for the holographic path is less then 7  and we can use the results of \fref{fig:TopBruteForce}. Since the corresponding effective region for the geodesic may be one or two sites larger then this the cases for sizes of $7$ and $8$ are shown in \fref{fig:ConcatBruteForce}b.

Using these results it can be seen that the path length for every transition between layers from bottom to top is $|\gamma_{A,\mathrm{part}}^{\mathrm{Holo}}| \leq 3|\gamma_{A,\mathrm{part}}^{\mathrm{Geo}}|$. Similarly comparing the geodesic and holographic path lengths for the top layer (effective holographic region less then 7) then again we find that $|\gamma_{A,\mathrm{part}}^{\mathrm{Holo}}| \leq 3|\gamma_{A,\mathrm{part}}^{\mathrm{Geo}}|$ and therefore we get the bound from \eref{eq:HolographicScreenBounds}. This bound is strict since when when moving from the boundary layer up to the the first layer the holographic path may never saturate the upper bound of \eref{eq:HolographicScreenBounds}.


This kind of relationship between maximal holographic screens and geodesic paths appears to be naturally generalisable to other MERA constructions including those in higher dimensions. The construction given to produce the maximal holographic screen places a natural upper bound on the geodesic length when we are working with a one dimensional boundary, for higher dimensions this extends to placing an upper bound on the minimal surface size. Finally it is worth commenting that while we expect this relationship between the minimal surface size and the maximal holographic screen size to hold in higher dimensions, there may be pathological cases where this breaks down. One example may be if the size of the boundary region in one dimension is significantly smaller then in the other, or if region is curved and the boundary region is no longer a convex shape.

\section{Details for the numerics}
\label{app:Numerics}

The family of Hamiltonians $H(k)$ of \eref{eq:AnyonicHam} is the anyonic analog of the standard spin $\frac{1}{2}$ antiferromagnetic Heisenberg model. This model arises from a deformation of the $SU(2)$ symmetry of the regular Heisenberg model, which restricts the possible spin values to half integers no greater then $\frac{k}{2}$. These deformed Heisenberg models exist as points, labelled by integers $k\geq 2$, along the XXZ spin chains:
\begin{eqnarray}
&\!\!\!\!\!\!\!\!\!\!\!\!\!\!\!\!\!\!\!\!\!\!\!\!\!\!\!\!\!\!
 H(k) = -\frac{1}{2 d(k)}\sum_{i} \Big[(\sigma^{x}_{i} \sigma^{x}_{i+1} + \sigma^{y}_{i}\sigma^{y}_{i+1})\nonumber
\\
&\quad  +\frac{d(k)}{2} \left(1 -\sigma^{z}_{i} \sigma^{z}_{i+1}\right) + i\sqrt{1-\frac{d^{2}(k)}{4}} (
\sigma^{z}_{i+1} - \sigma^{z}_{i}) \Big].
\label{eq:Hamiltonian}
\end{eqnarray}

As shown in Ref.~\cite{MCAnyonicEnergy}, and used in the main text, $H(k)$ can also be understood as a deformed Heisenberg model (for $k\geq 2$).

The ground state energy density is:
\begin{eqnarray}
\frac{E(k)}{N} &= \frac{d^{2}(k)-4}{4d(k)} \int_{-\infty}^{\infty} dx \frac{\sech(\pi x)}{\cosh\left[2 x \arccos\left(\frac{d(k)}{2}\right)\right] -\frac{d(k)}{2}}
\end{eqnarray}
where $d(k) = 2 \cos\left(\frac{\pi}{k+2}\right)$ and the central charge is 
\begin{equation}
c(k) = 1-\frac{6}{(k+1)(k+2)}. 
\label{eq:CentralCharge}
\end{equation}
While the total Hamiltonian is hermitian (up to boundary terms), the local interaction terms are non-hermitian. These local non-hermitian contributions are important as when they are removed $H(k)$ no longer corresponds to unitary minimal models with central charges defined by \eref{eq:CentralCharge}. Instead each Hamiltonian becomes a bosonic CFT with central charge $c = 1$. As mentioned in the main text these models are a deformed spin $\frac{1}{2}$ antiferromagnetic Heisenberg model:

\begin{equation}
H(k) = - \sum_{i} h_{i}, \qquad h_{i} = \vcenter{\hbox{\includegraphics[scale=0.4]{AnyonicHam}}},
\label{eq:AnyonicHamApp}
\end{equation}
where the $h_{i}$ term (depicted here in the anyon fusion basis) projects onto the state $\ket{\frac{1}{2}_{i}\times \frac{1}{2}_{i+1} \rightarrow 0}$, i.e. physically the projection onto the spin $0$ fusion space of two (deformed) spin $1/2$ particles at sites $i$ and $i+1$.

\begin{table}[t!]
{\small
\begin{tabular}{c|c|c|c|c|c}
$k = $ & $\Delta_{\mathrm{rel}} E(k)$ & $\Delta_{\mathrm{rel}} c(k)$ & $\Delta_{\mathrm{rel}} \left(\Delta^{(1)}\right)$ & $\Delta_{\mathrm{rel}} \left(\Delta^{(2)}\right)$ & $\Delta_{\mathrm{rel}} \left(\Delta^{(3)}\right)$
\\ \hline
2 & $2.3\times10^{-7}$ & $3.2\times10^{-3}$ & $1.2\times10^{-2}$ & $2.7\times10^{-3}$ & $7.9\times10^{-3}$
\\
3 & $4.4\times10^{-6}$ & $2.5\times10^{-2}$ & $3.2\times10^{-2}$ & $4.0\times10^{-2}$ & $3.4\times10^{-2}$
\\
4 & $1.6\times10^{-6}$ & $1.0\times10^{-2}$ & $1.2\times10^{-2}$ & $2.7\times10^{-3}$ & $7.9\times10^{-3}$
\\
5 & $9.4\times10^{-6}$ & $1.4\times10^{-2}$ & $^{-}3.1\times10^{-2}$ & $^{-}5.0\times10^{-3}$ & $1.4\times10^{-2}$
\\
6 & $1.0\times10^{-5}$ & $1.4\times10^{-2}$ & $4.1\times10^{-2}$ & $^{-}7.6\times10^{-3}$ & $2.0\times10^{-2}$
\\
7 & $1.1\times10^{-5}$ & $1.4\times10^{-2}$ & $^{-}7.2\times10^{-2}$ & $4.6\times10^{-4}$ & $5.2\times10^{-2}$
\\
8 & $1.1\times10^{-5}$ & $1.4\times10^{-2}$ & $^{-}1.1\times10^{-2}$ & $1.0\times10^{-3}$ & $7.1\times10^{-2}$
\\
9 & $1.2\times10^{-5}$ & $1.4\times10^{-2}$ & $4.3\times10^{-2}$ & $3.5\times10^{-3}$ & $^{-}6.6\times10^{-2}$
\\
10 & $1.2\times10^{-5}$ & $1.5\times10^{-2}$ & $2.0\times10^{-2}$ & $^{-}8.7\times10^{-4}$ & $^{-}6.5\times10^{-2}$
\\
$\infty$ & $1.3\times10^{-5}$ & $1.7\times10^{-2}$ & N/A & $6.0\times10^{-3}$ & N/A
\end{tabular}}
\caption{The relative errors in energy density $E(k)$, central charge $c(k)$ and the first two non-trivial scaling dimensions $\Delta^{(1,2)}$, for the numerical ground state calculations for the Hamiltonian given in Eq.~\ref{eq:Hamiltonian} and Eq.~\ref{eq:AnyonicHamApp}.}
\label{tab:ResultsAnyonicMERA}
\end{table}

The presence of these non-hermitian terms is also the reason that we cannot use the usual, non-symmetric, MERA for spin chain to simulate this family of models. Instead, we must resort to the anyonic version of the MERA \cite{anyonicMERA}. We studied the $k = 2,\cdots,10$ models and the $k=\infty$, the last of which corresponds to the standard spin $\frac{1}{2}$ antiferromagnetic Heisenberg model. We obtained the anyonic MERA representation of each ground state as described in Ref.~\cite{anyonicMERA}. In the simulations, we kept five transition layers and an additional scale-invariant layer. We assigned degeneracy 5 and 3 to irreps $\frac{1}{2}$ and $\frac{3}{2}$ respectively, on each bond index of the MERA, and zero degeneracy to all other irrep labels. (Except in the $k=2$ case, where spin $\frac{3}{2}$ does not exist.) The relative errors in ground state energy, central charge, and first couple (non-trivial) scaling dimensions are listed in table \ref{tab:ResultsAnyonicMERA}.

\begin{figure}[t]
\includegraphics[width=\columnwidth]{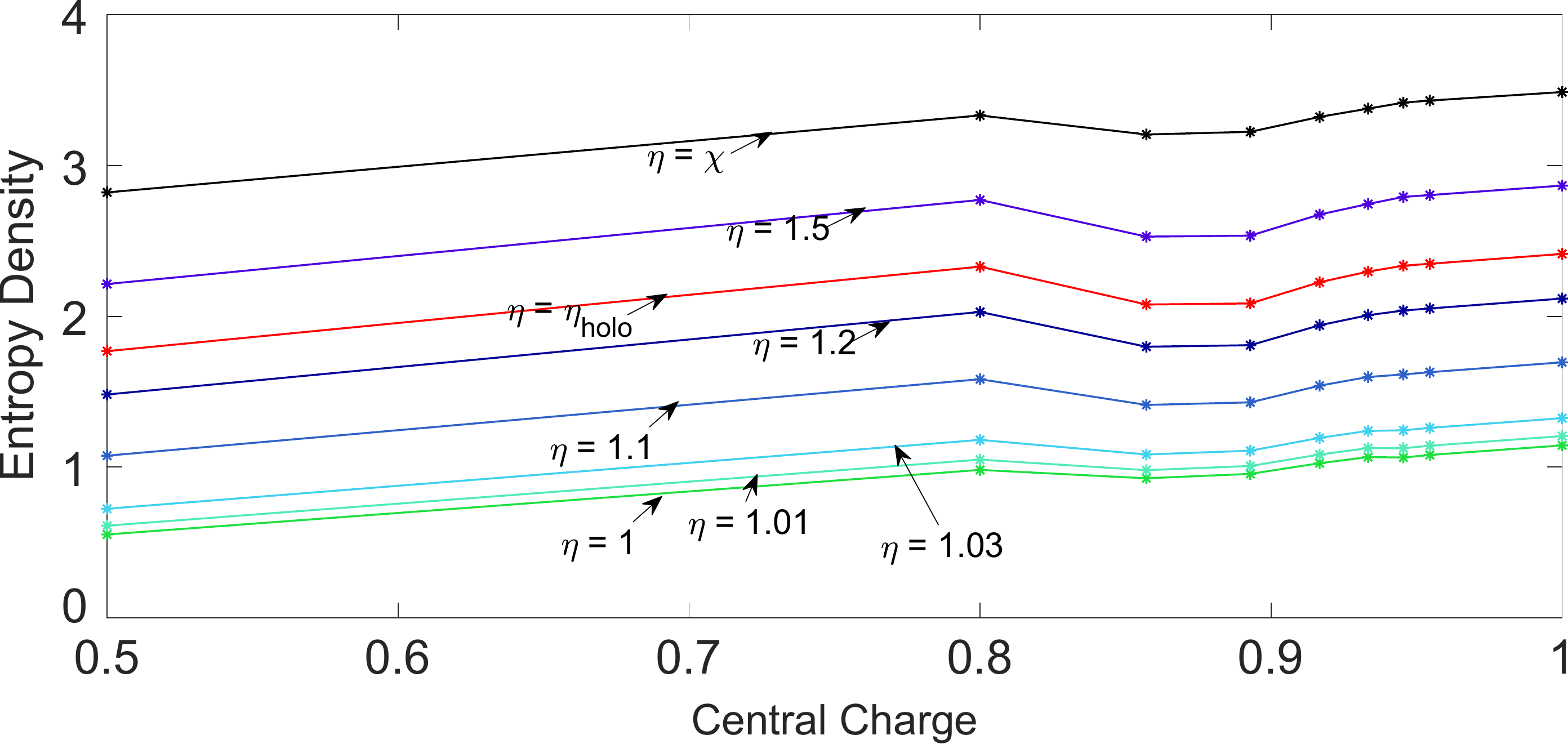}
\caption{
The second R\'enyi entropy density $S^{(2)}({\rho^{\mathrm{bk}}_{\gamma_{A}}}(\eta))/|\gamma_{A}|$ of bulk sites on the geodesic holographic screen $\gamma_{A}$ (see \fref{fig:holoscreen}) for ground states of the Hamiltonian $H(k)$ for various values of $\eta$ for each $k$, and $\eta_{\rm holo} \equiv \frac{\sqrt{2\chi}}{\sqrt{\chi+1}}$. The connections for different models for a given tuning parameter is not to indicate a predicted model that the authors propose, but rather to aid the reader in separating the different sets of lifting results based on the values of $\eta$ used.}
\label{fig:results_all}
\end{figure}

In addition, to the results presented in plot \fref{fig:NumericalResults}, for each value of $k$ we studied six additional values of tuning parameters---corresponding to six further dual bulk states for each critical ground state--- focusing on the $\eta \rightarrow 1$ limit. We considered $\eta = \{1,1.01,1.03,1.1,1.2,1.5\}$ where $\eta_{j}$ takes the same value in all cases, in addition we considered the holographic limit $\eta_{j} = \frac{\sqrt{2\chi_{j}}}{\sqrt{\chi_{j}+1}} \underset{\chi_{j}\rightarrow \infty}{\rightarrow} \sqrt{2}$ which is roughly between the $\eta = 1.2$ and $\eta = 1.5$ cases. We also considered the maximal entropy limit where $\eta_{j} = \chi_{j}$. In these last two values of $\eta$ we have set $\eta$ separately on each charge sector, $j$, based on the bond dimension, $\chi_{j}$, associated to said sector.

The results for all values of $k$ and $\eta$, excluding values for $k = 3$, are plotted in \fref{fig:results_all}. We restrict ourselves to regions $A$ between two spines (sequences of isometries connected via the middle sites) which eventually are neighbouring at some renormalisation scale. This means that the region $A$ corresponds to the boundary sites between the middle sites of the spine. This then means that the geodesic $\gamma_{A}$ corresponds to only paths which are of the form of \fref{fig:LeftBruteForce}(a) (see Appendix \ref{AppRefGeoCone}) and therefore is both the maximal holographic screens and the graph geodesic paths.

\begin{figure}[t]
\includegraphics[width=0.64\columnwidth]{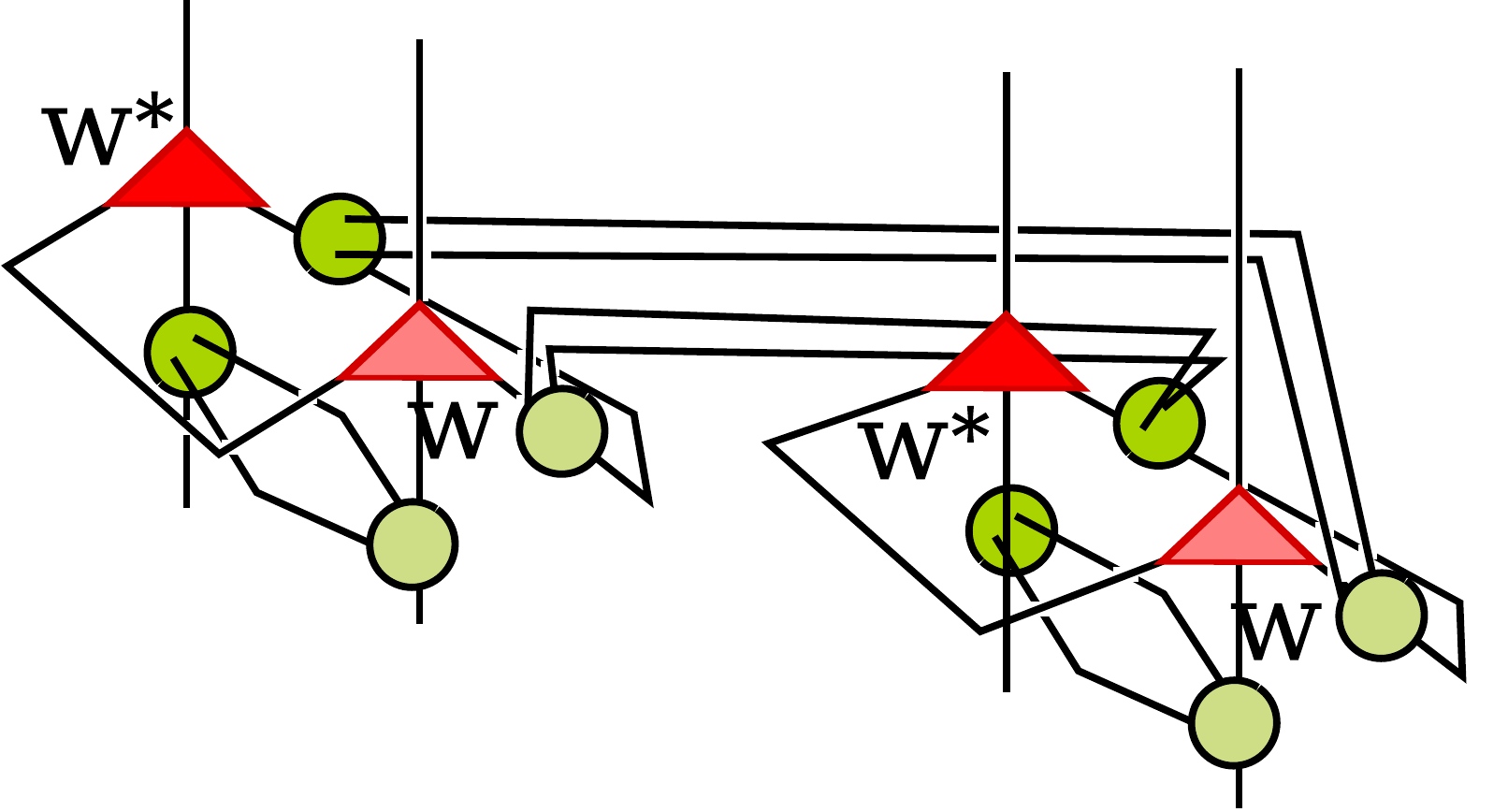}
\caption{The transfer matrix that was used to compute the entropy density.
}
\label{fig:Transfer_Matrix}
\end{figure}

To compute the entropy density of this we compute the eigenvalues $\{\lambda_{i}\}_{i = 0, \cdots}$ of a transfer matrix shown in figure \fref{fig:Transfer_Matrix}. By doing this we have computed the asymptotic entropy density $S^{(2)}(\rho^{\rm bk}_{\gamma_{A}})/|\gamma_{A}| = \log_{2}(\lambda_{1})$ as $|A| \rightarrow \infty$ and so only the scale invariant layer contributes to this calculation. We exclude the $k = 3$ values due to lack of convergence in the entropy density results. In all these cases we see a complete ordering of the entropy density as defined by the tuning parameter $\eta$ just as we would expect in the non-symmetric case. A partial order may be able to be imposed in the symmetric case by tuning the $\eta_{j}$ values separately for each $j$ so that we may have $\eta^{(1)}_{j} < \eta^{(2)}_{j}$ but $\eta^{(1)}_{j^{'}} > \eta^{(2)}_{j^{'}}$.

Given this observation we can now consider the two limits of $\eta$ and compare them to the non-symmetric case. The first limit of $\eta = \chi$, which for the symmetric case becomes $\eta_{j} = \chi_{j}$, we expect the entropy density to be the maximal possible entropy density of $\log(\chi)$ for the non-symmetric case. In the symmetric case the analogue would be expected to be $\log(\sum_{j} \chi_{j} d_{j})$ where $d_{j}$ is the quantum dimension of the charge $j$ which is defined for $SU(2)_{k}$ as:

\begin{equation}
d_{j} = \frac{\sin\left( \frac{(2j+1)\pi}{k+2} \right)}{\sin\left( \frac{\pi}{k+2} \right)}
\end{equation}

In table \ref{tab:Results_Eta_Chi} we compare the computed entropies for when $\eta_{j} = \chi_{j}$ for all values of $k$. Based on the non-symmetric results, where the state should be a maximally mixed state we expect this to go as $S(I_{\gamma_{A}})/|\gamma_{A}| = -\log_{2}\left(\frac{\sum_{j} \chi_{j} d_{j}}{(\sum_{j} \chi_{j} d_{j})^{2}}\right)$ (using base 2 for all entropy calculations) where $j$ is summed over all possible charge labels. We find that the entropy for the symmetric models does not completely saturate this bound as there is additional information encoded in the probability of existing in the different gauge DOFs. As confirmation that this is the case we see that when $k=2$ and we have only stored a single gauge degree of freedom, this bound perfectly predicts the entropy density.

\begin{table}[t!]
\begin{tabular}{c|c|c|c|c}
$k = $ & $S(\widetilde{\rho}_{\gamma_{A}})/|\gamma_{A}|$ & $S(I_{\gamma_{A}})/|\gamma_{A}|$ & $\Delta S/|\gamma_{A}|$ & $\Delta_{\mathrm{Rel}} S/|\gamma_{A}|$
\\ \hline
2 & $2.8219$ & $2.8219$ & $0$ & $0$
\\
4 & $3.7925$ & $3.3326$ & $0.4599$ & $0.1213$
\\
5 & $3.9773$ & $3.2063$ & $0.7710$ & $0.1939$
\\
6 & $4.0941$ & $3.2240$ & $0.8701$ & $0.2125$
\\
7 & $4.1727$ & $3.3231$ & $0.8496$ & $0.2036$
\\
8 & $4.2283$ & $3.3777$ & $0.8506$ & $0.2012$
\\
9 & $4.2691$ & $3.4177$ & $0.8514$ & $0.1994$
\\
10 & $4.2999$ & $3.4313$ & $0.8686$ & $0.2020$
\\
$\infty$ & $4.4594$ & $3.4876$ & $0.9719$ & $0.2179$
\end{tabular}
\caption{Entropy densities for computed results for $\eta_{j} = \chi_{j}$, $S(\widetilde{\rho}_{\gamma_{A}}^{\mathrm{bk}}$, and the expected maximal entropy density, $S(I_{\gamma_{A}})/|\gamma_{A}|$. In addition to this the differences between the computed and predicted entropy densities, $\Delta S$, are given as well as the relative difference $\Delta_{\mathrm{Rel}} S$. 
}
\label{tab:Results_Eta_Chi}
\end{table}

We also wish to study the other limit, where $\eta_{j} = 1$, and compare it with the prediction of describing the entropy of the original CFT theory. For Renyi-2 entropy the relationship between the entropy of the CFT on region $A$, $\rho^{\mathrm{CFT}}_{A}$ and the central charge $c$ and the subsystem size $|A|$, is:

\begin{equation}
S^{(2)}(\rho_{A}^{\mathrm{CFT}}) = \frac{c}{4} \log_{2}(|A|).
\end{equation}
For the path that we are considering there is a relationship between $|A|$ and the path length through the bulk $|\gamma_{A}|=2N$ given by:

\begin{equation}
|A| = 2\frac{3^{N}-1}{3-1} \Rightarrow |\gamma_{A}| = 2\log_{3}\left(|A|+1\right)
\end{equation}
This means that we expect our entropy density to take the form:

\begin{equation}\label{eq:Predicted_Linear}
S^{(2)}(\rho_{A}^{\mathrm{CFT}})/|\gamma_{A}| \approx \frac{c}{8} \log_{2}(3) \approx 0.20 c.
\end{equation}

To study this first we want to perform linear regression on our $\eta = 1$ limit to compare to the prediction from \eref{eq:Predicted_Linear} for a linear regression to $S^{(2)}(\rho_{A}^{\mathrm{CFT}})/|\gamma_{A}| = m c + x^{(0)}$. We then compare the regression to linear, quadratic and cubic models and consider the t-statistic of the coefficients of these models to determine the most appropriate model. The t-statistic in this case is a measure of the probability that the results (the entropy density) are completely uncorrelated to the other variables (the central charge). In particular this probability is the chance that a random Gaussian distribution could give rise to this distribution. The t-statistic has an associated p-value which is interpreted as the probability of having sampled random points to generate a correlation of this magnitude or greater. Further we work out the t-statistics for comparing the higher order models to the linear model, in this case we repeat the analysis on the difference between the entropy density an the linear model.

Doing this we get $m = 1.131 \pm 0.095$ with a t-statistic of 11.9 (p value of $6.8\times 10^{-6}$) and $x^{(0)} = -0.003\pm 0.083$ with a t-statistic of -0.0384 (p value of $0.97$). This states that the disturbance of the intercept away from the theoretically predicted value of $0$ is extremely likely ($\sim97\%$) to be due to random noise (here due to the fact that the MERA with finite bound dimension is only an approximation to the ground state). The linear contribution is highly unlikely to be due to random noise ($<10^{-3}\%$). Further the value we find is that the slope is about 5
 times greater then we predicted and this is attributed to be due to additional decoherence that occurs due to the coupling of the charge sectors to the bulk degrees of freedom.

\begin{table}[t!]
\begin{tabular}{|c|c|c|c|c|}
\hline
$x^{n}$ & Coefficient & Standard Error & t Stat & p Value
\\ \hline
\multicolumn{5}{c}{Linear, Null = Gaussian Noise}
\\ \hline 
$x^{0}$ & $-0.003$ & $0.083$ & $-0.038$ & $0.97$
\\
$x^{1}$ & $1.131$ & $0.095$ & $11.89$ & $6.8\times10^{-6}$
\\
\hline
\multicolumn{5}{c}{Quadratic, Null = Gaussian Noise}
\\ \hline 
$x^{0}$ & $-0.10$ & $0.39$ & $-0.26$ & $0.81$
\\
$x^{1}$ & $1.4$ & $1.1$ & $1.27$ & $0.25$
\\
$x^{2}$ & $-0.19$ & $0.74$ & $-0.26$ & $0.81$
\\
\hline
\multicolumn{5}{c}{Cubic, Null = Gaussian Noise}
\\ \hline 
$x^{0}$ & $-7.1$ & $3.3$ & $-2.1$ & $0.086$
\\
$x^{1}$ & $31$ & $14$ & $2.2$ & $0.078$
\\
$x^{2}$ & $-39$ & $19$ & $-2.1$ & $0.087$
\\
$x^{3}$ & $16.9$ & $8.0$ & $2.1$ & $0.088$
\\
\hline
\hline
\multicolumn{5}{c}{Quadratic, Null = Linear + Gaussian Noise}
\\ \hline 
$x^{0}$ & $-0.10$ & $0.39$ & $-0.25$ & $0.81$
\\
$x^{1}$ & $0.3$ & $1.1$ & $0.25$ & $0.81$
\\
$x^{2}$ & $-0.19$ & $0.74$ & $-0.26$ & $0.81$
\\
\hline
\multicolumn{5}{c}{Cubic, Null = Linear + Gaussian Noise}
\\ \hline 
$x^{0}$ & $-7.1$ & $3.3$ & $-2.1$ & $0.086$
\\
$x^{1}$ & $30$ & $14$ & $2.1$ & $0.078$
\\
$x^{2}$ & $-39$ & $19$ & $-2.1$ & $0.087$
\\
$x^{3}$ & $16.9$ & $8.0$ & $2.1$ & $0.088$
\\ \hline
\end{tabular}
\caption{Table of coefficients for polynomial fits to the $\eta = 1$ data. Along with the fits the standard error, t-statistic and associated p-Value generated by the regression are included in this table. Analysis of these results indicates that the linear model is the most statistically significant with a probability of the distribution being random of about $7\times 10^{-6}$. All other models have cofficients  with significantly larger probabilities of the distribution being random, these at least on the order of several parts in a hundred.
}
\label{tab:T_Stats}
\end{table}

For completeness we also study the likelihood of this data being described by a higher order polynomial in the central charge. This was done by computing the t-statisitics for the coeficients when performing linear regression to quadratic and cubic models, in these cases we find that the best fit to a quadratic model is highly likely to be generated purely by the gaussian noise that is assumed during linear regression. For the cubic model each coefficient has a $\sim 8\%$ chance of being random noise, this is better then the quadratic model but in the linear model the probability of the linear relation being due to random noise is roughly a factor of 1000 smaller then any polynomial relationship in the cubic model. For completeness we also compare this to the linear model of $S^{(2)}(\rho_{A}^{\mathrm{CFT}})/|\gamma_{A}| = 1.131 C - 0.003$ and find similar results (where the null model is gaussian noise plus the linear model). These results are given in table \ref{tab:T_Stats}.

At the extreme points of $\eta$ ($\eta_{j} = 1$ and $\eta_{j} = \chi_{j}$) we see there is an almost linear relationship between the entropy density and the central charge.  Between these extreme points we see a deviation from linearity which becomes most extreme around $\eta = \eta_{\mathrm{Holo}}$. These deviations occur most significantly for $k = 4,5,6$. There are two possible reasons that this may be the case, the first is that we should use a different tuning parameter for the different charge sectors. However the behaviour of constant tuning parameter $\eta_{j} = 1.2$ and a the holographic tuning $\eta_{j} = \sqrt{2\chi_{j}}/\sqrt{\chi_{j}+1}$ which is roughly $\eta = (1.29,1.22)$ for the two relevant charge sectors. This suggests that the behaviour has to do with absolute values (or absolute values relative to the charge sector bond dimension) as opposed to due to relative tunings between charge sectors.

The abnormal behaviour could also arise from the fact that as $\eta$ varies between $1$ and $\chi$ we get a mixture of behaviours of the lifting procedure. Drawing from our intuition of the non-symmetric case, in the $\eta = 1$ case we find that the state along the maximal holographic screen should behave as a section of the boundary CFT. On the other hand, in the $\eta = \chi$ limit this gives rise to a maximally mixed state. In between these limits we expect there to be both mixed state contributions as well as the pure state contribution. Notwithstanding, since the mixing of $\alpha$ and $\beta$ contributions to the lifting tensor occurs not only at the holographic screen, but also above it, this suggest that non-linear feedback as we transition between the $\eta = 1$ and $\eta = \chi$ extremes could be important. For this reason the anomalous behaviour around $\eta = \eta_{\mathrm{Holo}}$ may be indicative of some kind of transition from CFT like behaviour to a mixed state like behaviour. This non-linearity with respect to the central charge may be the true behaviour around these parameters, however the authors suspect this behaviour may just be the lifting procedure exacerbates the numerical instabilities from the original MERA procedure. For this reason it may be worthwhile exploring this for a possible transition and any associated order parameters in future work.

\end{document}